\documentclass{emulateapj}
\usepackage{apjfonts}

\def\sfrac#1#2{\, \hbox{\raisebox{0.5ex}{\scriptsize #1}} \!/\! \hbox{\raisebox{-0.5ex}{\scriptsize #2}} \,}

\shorttitle{\sc Quenched Cold Accretion}
\shortauthors{\sc Churchill {\etal}}

\input{iondefs.sty}
\begin{document}

\title{Quenched Cold Accretion of a Large Scale Metal-Poor Filament due to \\
Virial Shocking in the Halo of a Massive $\lowercase{z}=0.7$ Galaxy}

\author{\sc
Christopher W. Churchill\altaffilmark{1,2},
Glenn G. Kacprzak\altaffilmark{3,4}, 
Charles C. Steidel\altaffilmark{5},
Lee R. Spitler\altaffilmark{3}, \\
Jon Holtzman\altaffilmark{1},
Nikole M. Nielsen\altaffilmark{1},
and
Sebastian Trujillo-Gomez\altaffilmark{1}
}

\altaffiltext{1}{New Mexico State University, MSC 4500, Las Cruces, NM 88003,
USA}

\altaffiltext{2}{Visiting Professor, Swinburne University of
Technology, Victoria 3122, Australia}

\altaffiltext{3}{Swinburne University of Technology, P.O. Box 218,
Victoria 3122, Australia}

\altaffiltext{4}{Australian Research Council Super Science Fellow}

\altaffiltext{5}{California Institute of Technology, MS 105-24,
Pasadena, CA 91125, USA}

\begin{abstract}

Using {\it HST}/COS/STIS and HIRES/Keck high-resolution spectra, we
have studied a remarkable {\HI} absorbing complex at $z=0.672$ toward
the quasar Q1317+277.  The {\HI} absorption has a velocity spread of
$\Delta v = 1600$~{\kms}, comprises 21 Voigt profile components, and
resides at an impact parameter of $D=58$ kpc from a bright, high mass
($\log M_{\rm vir}/M_{\odot} \simeq 13.7$) elliptical galaxy that is
deduced to have a 6 Gyr old, solar metallicity stellar population.
Ionization models suggest the majority of the structure is cold gas
surrounding a shock heated cloud that is kinematically adjacent to a
multi-phase group of clouds with detected {\CIII}, {\CIV} and {\OVI}
absorption, suggestive of a conductive interface near the shock.  The
deduced metallicities are consistent with the moderate {\it in situ\/}
enrichment relative to the levels observed in the $z\sim 3$ {\Lya}
forest. We interpret the {\HI} complex as a metal-poor filamentary
structure being shock heated as it accretes into the halo of the
galaxy.  The data support the scenario of an early formation period
($z > 4$) in which the galaxy was presumably fed by cold-mode gas
accretion that was later quenched via virial shocking by the hot halo
such that, by intermediate redshift, the cold filamentary accreting
gas is continuing to be disrupted by shock heating.  Thus, continued
filamentary accretion is being mixed into the hot halo, indicating
that the star formation of the galaxy will likely remain quenched.  To
date, the galaxy and the {\HI} absorption complex provide some of the
most compelling observational data supporting the theoretical picture
in which accretion is virial shocked in the hot coronal halos of high
mass galaxies.

\end{abstract}

\keywords{quasars: absorption lines}

\section{Introduction}
\label{sec:intro}

It is well accepted that galaxies are intimately linked to the gaseous
cosmic web and that the evolution of galaxies is governed in large
part by the dissipative response of baryonic gas due to the trade off
of cooling and dynamical timescales as it accretes into dark matter
halos \citep[e.g.,][]{binney77,rees77,silk77,white78}. In general,
distinct modes of accretion are believed to operate and the mode is
dependent primarily upon the dark matter halo mass
\citep[e.g.,][]{birnboim03,dekel06}, with some dependence on
environment \citep[e.g.,][]{keres05} and on feedback
\citep[e.g.,][]{vandevoort11}.

``Hot-mode'' accretion is the mechanism in which inflowing gas is
shock heated as it is compressed by the hot hydrostatic gas halo.
This mode dominates around high mass galaxies, where the dynamical
time is shorter than the cooling time.  For the most part, the gas
accretes into the halo, but not necessarily onto the galaxy itself.
``Cold-mode'' accretion is the mechanism primarily around low mass
galaxies, where the cooling time is shorter than the dynamical time so
that hot hydrostatic halos do not form and cold gas accretes directly
onto the galaxy.  At lower redshifts, as densities decrease, the
cooling time is generally longer than dynamical time and the rate of
infalling material decreases so that cold mode accretion is a minor
channel of accretion onto galaxies \citep[e.g.,][]{keres05,dekel06},
though it remains an important channel for the growth of galaxies at
all redshifts \citep[e.g.,][]{vandevoort11}.

A possible third mode is the case in which cold dense filaments can
penetrate directly into a hot halo of a massive galaxy.  If the
conditions are favorable for short cooling times in the filament, then
shock heating can be avoided and the filamentary gas can directly
accrete onto the galaxy \citep[e.g.,][]{dekel06,vandevoort+scahye11}.
However, though cold streams can penetrate the hot atmospheres of
massive halos at $z\geq2$, this process significantly diminishes at
lower redshift \citep[e.g.,][]{keres09,faucher-gigure11}, though there
is plausible evidence of this process occuring in some $z<1$ galaxies
\citep{kcems,ribaudo11,thom11,ggk-q1317}.

One observable signature of a filament may be a ``complex'' of
{\HI} absorption with a large velocity spread ($\Delta v > 1000$
{\kms}).  Alternatively, such {\HI} absorbing complexes may trace the
warm hot intergalactic medium (WHIM), or the intracluster and/or
intragroup medium.  If an observed {\HI} absorption complex arises
from a filament, it is plausible that the filament may be accreting
from the intergalactic medium into a galaxy halo or even directly onto
a galaxy.  Thus, {\HI} absorption complexes provide unique
astrophysical laboratories for placing constraints on our
understanding of the intergalactic medium, and the processes giving
rise to extended galaxy halos in the context of dark matter
overdensities.

At high redshifts ($z>2$), kinematically extended {\HI} absorbing
complexes were studied by \citet{cowie96} with Keck/HIRES spectra.
They deduced that these structures were filamentary in nature.  At
lower redshift ($z<0.22$) \citet{shull98}, \citet{tripp01},
\citet{shull03}, and \citet{aracil06a} studied three different {\HI}
complexes with high resolution ultraviolet spectra in which metal
lines were detected.  For these three complexes, several galaxies were
found in the vicinity, suggesting moderate size groups.  Each of these
studies favored a different scenario of explanation for the physical
picture of the {\HI} complex, including {\OVI} arising in
``nearside/backside'' shocked infall into the potential well of the
galaxy group \citep{shull03}, intragroup gas or an unvirialized
filamentary structure through the group \citep{tripp01}, and tidally
stripped material from one of the nearby galaxies \citep{aracil06a}.

The spectrum of the quasar Q1317+277 (TON 153, CSO 0873,
J131956+272808, $V=16.0$, $z_{\rm em} = 1.017$) exhibits a dramatic
$\Delta v > 1000$~{\kms} {\HI} complex observed in {\Lya} absorption
at $z=0.672$.  Several optical and ultraviolet spectra of the quasar
are available and have been the focus of this {\HI} complex and/or the
Lyman limit metal-line system at $z=0.660$
\citep{ss92,cat1,cat2,archiveI,ding05,paper1,kcems,cwc-anatomy,ggk-q1317}.
Besides being at intermediate redshift, what is unique to this {\HI}
complex is that it lies at 58 kpc projected from a single bright
elliptical galaxy at $z_{\rm gal}=0.6719$ \citep{paper1}.

An analysis of the $z = 0.672$ {\HI} complex was presented in
\citet[][hereafter Paper I]{paper1} based upon {\it HST\/} G160L/G190H
FOS spectra (PID 2424, PI: J. N. Bahcall), an {\it HST\/} E230M/STIS
spectrum (PID 8672, PI: Churchill), and a Keck/HIRES spectrum of the
quasar \citep[see][]{cwc-thesis}.  In the FOS spectrum, the {\HI}
complex was found to comprise five components of optically thin {\HI}
absorbing gas, which span a velocity range of $\Delta v =
1400$~{\kms}.  No metal-lines were clearly detected in the FOS, STIS,
and HIRES spectra.

Galaxies at the absorber redshifts were first reported by
\citet*{sdp94} as part of the their {\MgII} absorption-selected galaxy
survey.  The quantified morphological and spectral properties of these
galaxies were presented in Paper~I\nocite{paper1} based upon an
{\it HST}/WFPC2 F702W image (PID 5984; PI: Steidel) of the quasar
field and Keck/LRIS spectra of the galaxies.  Updated analysis of the
galaxies is included in \citet{kcems}, \citet{cwc-anatomy}, and
\citet{ggk-q1317}.  In summary, the galaxy at the redshift of the
{\HI} complex has $z_{\rm gal} = 0.6719$ and an impact parameter of
$D=58$ kpc.  It is classified as a late-type E/S0 galaxy.  The galaxy
at the redshift of the $z= 0.660$ absorber, which is a rich metal-line
Lyman limit system, has $z_{\rm gal} = 0.6610$, an impact parameter of
$D=104$, and is an inclined Sab galaxy.

In Paper~I\nocite{paper1}, we deduced that the gas in the {\HI}
complex has column densities in the range $14.5 \leq \log N({\HI})
\leq 15.5$, temperatures in the range $5.0 \leq \log T \leq 5.5$, and
upper limits on metallicity in the range $-1.0 \leq \log Z/Z_{\odot}
\leq -3.0$.  We further deduced the complex is consistent with a
combination of photo and collisional ionized gas.  Based upon
expectations of simulations \citep[e.g.][]{dave99}, we favored a shock
heated structure with chemical enrichment consistent with the high
redshift intergalactic medium.  Our interpretation was that the {\HI}
complex is a shock heated filamentary structure that originated as a
photoionized diffuse phase of gas at high redshift and is accreting in
the vicinity of the $z_{\rm gal} = 0.6719$ elliptical galaxy.

However, the FOS data did not provide the information required to
examine the detailed structure of the kinematic and ionization
conditions.  Thus, we were unable to constrain the kinematic
relationships between the putative shock heated gas, the photo and/or
collisionally ionized gas, and the $z_{\rm gal} = 0.6719$ galaxy.
Such information is critical for examining the physics of gas
accretion onto galaxies and for comparing with other observations and
with cosmological simulations.  For example, \citet{cowie96} claim
that clustered {\Lya} lines at $z=3$ are observed to have higher
ionization conditions at the velocity extremes and suggest this
layered structure is a signature of collapsing structures.  While this
may be a signature for intergalactic filaments, it is not clear that
accretion of a filament into a galaxy potential will yield the same
velocity-ionization structure.  In theoretical treatment
\citep[e.g.,][]{birnboim03,dekel06,birnboim07} and cosmological
simulations
\citep[e.g.,][]{keres05,keres09,vandevoort11,vandevoort+scahye11}
accretion onto massive galaxies is expected to shock heat upon entry
into a hot coronal halo, and the resulting kinematic-ionization
structure may reflect this very different process.

In order to better study the Q1317+277 {\HI} complex at $z=0.672$, we
have obtained a high resolution $R=18,000$ {\it HST}/COS spectrum of
the quasar, with focus on the {\HI} Lyman series lines and {\OVIdblt}
absorption.  We further improved our knowledge of the star formation
history, age, and mass of the $z_{\rm gal} = 0.6719$ galaxy by
obtaining multi-band imaging of the quasar field.  Our motivations
include (1) a thorough examination of the kinematic, ionization, and
chemical structure of the {\HI} complex, (2) a direct comparison with
the galaxy properties to help place the galaxy-absorber system in the
context of galaxy evolution scenarios predicted by theories and
cosmological simulations for an improved interpretation, and (3) 
examination whether there is a connection between the $z=0.672$ {\HI}
complex and the galaxy-absorber pair at $z=0.661$, perhaps in the form
of ``bridge'' of weak {\Lya} absorption between the galaxies.

The paper is structured as follows: Reduction and analysis of the
imaging and spectroscopic data are presented in \S~\ref{sec:data}.
Our analysis of the galaxy images is presented in \S~\ref{sec:imanal}
and our analysis of the absorption line data is presented in
\S~\ref{sec:specanal}.  In \S~\ref{sec:models} and
\S~\ref{sec:gasconditions}, we present our ionization modeling and
resulting constraints on the physical conditions of the absorbing gas.
We discuss and interpret the data and our findings in
\S~\ref{sec:discuss}.  We provide brief a conclusion in
\S~\ref{sec:conclude}.  Throughout this work, we assume the
cosmological parameters $\Omega_\Lambda=0.7$, $\Omega_m=0.3$,
$\Omega_k=0$, and $h = H_0/(100~\hbox{\kms}~{\rm Mpc}^{-1} )= 0.7$
\citep[based on][]{wmap}.

\begin{figure*}[bht]
\epsscale{0.95}
\plotone{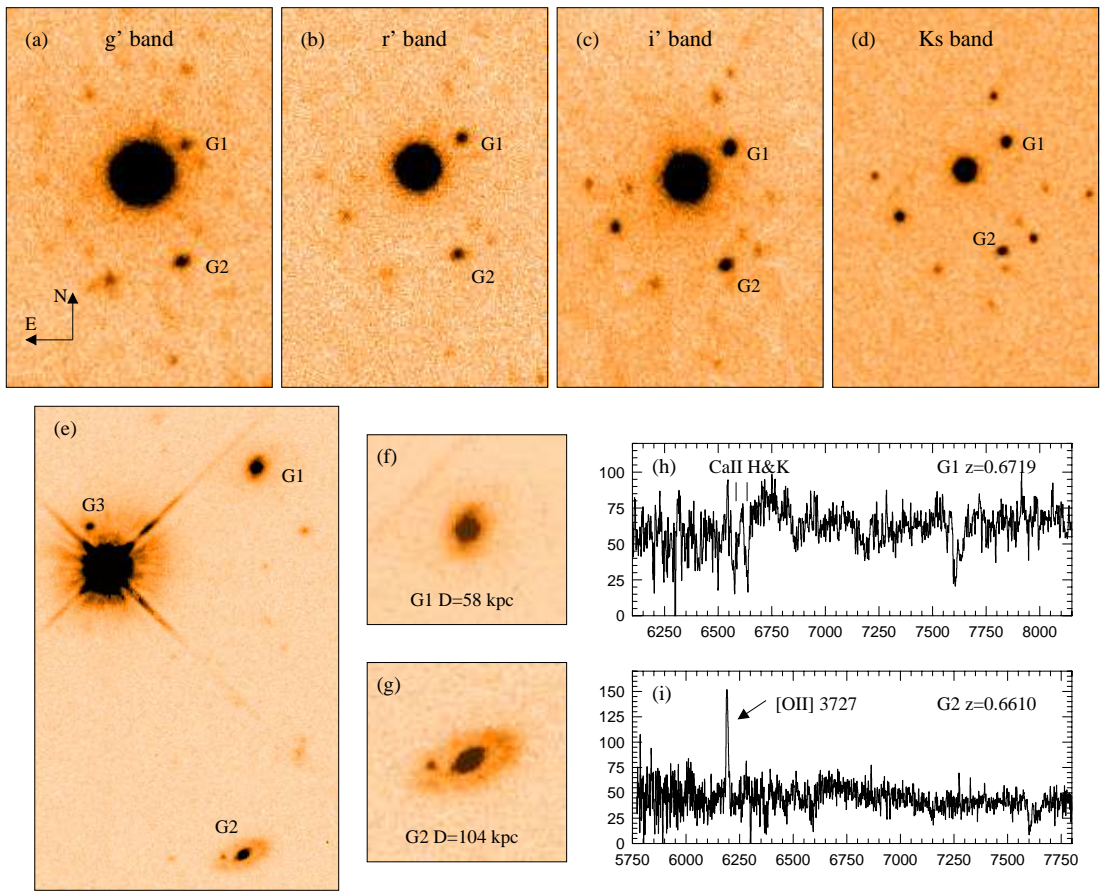}
\caption{Images of the quasar field and spectra of the absorbing
galaxies. --- ($a$--$c$) The $43{\arcsec} \times 60{\arcsec}$ sections
of the ground-based APO/SPIcam images centered on the quasar through
the SDSS $g'$, $r'$, $i'$ filters.  Galaxies G1 and G2 are identified.
--- ($d$) A $43{\arcsec} \times 60{\arcsec}$ section of the KPNO/IRAM
NICMOS image through the $K_s$ filter.  --- ($e$) A $12{\arcsec}
\times 22{\arcsec}$ section of the {\it HST}/WFPC2 image. Galaxies
G1 and G2 are labeled.  Also identified and labeled is object G3, a
compact galaxy (with unknown redshift) at angular separation
$2.1{\arcsec}$ in the north by north-east direction from the quasar.
--- ($f$) A $4{\arcsec} \times 4{\arcsec}$ view of galaxy G1 at $D=58$
kpc from the quasar. --- ($g$) A $4{\arcsec} \times 4{\arcsec}$ view
of galaxy G2 at $D=104$ kpc from the quasar.  --- ($h$) The Keck/LRIS
spectrum (counts versus wavelength) of G1, yielded $z_{\rm
gal}=0.6719$ from the {\CaII} H\&K doublet.  --- ($i$) The Keck/LRIS
spectrum (counts versus wavelength) of G2, yielded $z_{\rm
gal}=0.6610$ from the {\OII} $\lambda 3727$ emission line.}
\label{fig:data}
\vglue 0.1in
\end{figure*}

\section{Observations, Reductions, and Calibrations}
\label{sec:data}

We have acquired new data on the $z=0.660$ and $z=0.672$ absorption
and the two galaxies G1 and G2.  First, we obtained a cycle-17
$R\simeq 18,000$ COS spectrum (PID 11667; PI: Churchill) of the quasar
covering the transitions examined in the FOS spectra.  Our goals
include measuring higher detection sensitivities and detailed
kinematics of the absorption lines.  Second, we obtained $g'$, $r'$,
$i'$, and $K_s$ ground-based images of the quasar field.  The
multi-band images provide colors from which galaxy stellar
populations, metallicities, and masses can be estimated.

\subsection{Ground-based Images}

As part of a larger campaign for imaging {\MgII} absorption galaxies,
we obtained $g'$, $r'$, $i'$ band images of the quasar field using the
Seaver Prototype Imaging camera (SPIcam) on the Apache Point
Observatory's (APO) 3.5-m telescope.  The detector is a $2048 \times
2048$ CCD with 24~$\mu$m pixels, giving an unbinned plate scale of
0.14{\arcsec} pixel$^{-1}$ and a field of view of $4.78{\arcmin}
\times 4.78{\arcmin}$.  We binned the CCD $2 \times 2$ during readout
for a plate scale of 0.28{\arcsec} pixel$^{-1}$.  Multiple $g'$, $r'$,
and $i'$ band images were obtained on the nights of 2006 March 24-25,
for which the seeing varied between 1.2-1.5{\arcsec}, and additional
$g'$ band images were obtained the night of 2007 March 15 for which
the seeing was 0.7{\arcsec}.  The total summed exposure times are
5190, 4630, and 4350 seconds for the $g'$, $r'$, and $i'$ filters,
respectively.

Each frame was reduced using standard IDL and IRAF\footnote{IRAF is
written and supported by the IRAF programming group at the National
Optical Astronomy Observatories (NOAO) in Tucson, Arizona. NOAO is
operated by the Association of Universities for Research in Astronomy
(AURA), Inc.\ under cooperative agreement with the National Science
Foundation.} scripts and tasks.  Flat fielding incorporated a
combination of dome and sky flats.  Cosmic rays were removed in each
individual frame.  The astrometry was calibrated by position matching
of USNO A2.0 stars in the field.  Final images were obtained
by coadding the individual calibrated frames.

The photometric zero points were obtained using stars in SDSS images.
Color terms are required because the filter plus detector throughput
of the APO facility is not identical to that of the SDSS facility.
These color terms, which are of order 0.1, were determined from SPIcam
and SDSS images of roughly 30 quasar fields from our more extensive
database.

The near-infrared band images were obtained on 1994 February 24 using
the Kitt Peak Mayall 4-m telescope through the $K_s$ filter
(1.99--2.32~$\mu$m) with the IRIM NICMOS III $256\times256$ array
camera.  These images were obtained as part of the campaign
culminating in the work of \citet{sdp94}.  The field of view is
$154\arcsec \times 154\arcsec$ with plate scale of $0.6\arcsec$
pixel$^{-1}$.  The NICMOS images were reduced using the contributed
IRAF package DIMSUM\footnote{{\it
http://iraf.noao.edu/iraf/ftp/contrib/dimsumV3/.} DIMSUM was
contributed by P. Eisenhardt, M. Dickinson, S. A. Stanford, \&
F. Valdez.}.  Updated photometric zero points were determined using
stars from the 2MASS point-source catalog \citep{strutskie06}.

All photometric measurements were conducted using SExtractor
\citep{bertin96}; we adopted the AUTOMAG results.  The dust maps of
\citet{schlegel98} were used to correct for Galactic dust extinction.

In Figures~\ref{fig:data}$a$-\ref{fig:data}$d$, we present
$43{\arcsec} \times 60{\arcsec}$ sections of the ground-based images
centered on the quasar Q1317+277.  A $12\arcsec \times 22\arcsec$
section of the {\it HST}/WFPC2 F702W image is presented as
Figure~\ref{fig:data}$e$. Galaxies G1 and G2 labeled in all images.
In the {\it HST\/} image, note the object to the north by north-east
within 2.1{\arcsec} of the quasar.  We have no estimate of the
redshift of this object, which we label G3.  Reduction and analysis of
the F702W image was described in Paper~I\nocite{paper1},
\citet{kcems}, and \citet{cwc-anatomy}.  Expanded views of the two
galaxies are presented in Figures~\ref{fig:data}$f$ and
\ref{fig:data}$g$.  The Keck/LRIS spectra of galaxies G1 and G2
(originally described in Paper~I)\nocite{paper1} are presented in
Figures~\ref{fig:data}$h$ and \ref{fig:data}$i$.  The G1 galaxy
redshift was determined using Gaussian centroiding to the {\CaII}
absorption features and the redshift of G2 was determined using
Gaussian centroiding to the {\OII} $\lambda 3727$ emission line.

\subsection{{\it HST}/COS spectrum}

We obtained a cycle-17 $R\simeq 18,000$ COS spectrum (PID 11667; PI:
Churchill) of the quasar covering the transitions first examined in
the FOS spectra.  Two NUV/G185M spectra were obtained 2010 May 26 and
optimally co-added.  The first was centered at 1921~{\AA} for a 5420
sec exposure, and the second was centered at 1941~{\AA} for a 4970 sec
exposure.  The overlap region was 2223--2037~{\AA} on Stripe C, which
provided a total of 10,390 sec of integration on the the {\HI}
absorption complex.  The FUV/G160M spectrum was obtained 2010 June 26
centered at 1600~{\AA} for an 12,580 sec exposure.  

The spectra were reduced following the procedures outlined by
\citet{cos-dhb}.  We continuum fit the lower order shape of the
spectrum using the IRAF {\it sfit\/} task and refined the higher order
continuum features using our own code Fitter \citep{archiveI}.

For the $z=0.660$ and $z=0.672$ absorbers, the {\Lyb} and {\Lyg}
absorption lines were captured in the NUV on Segment A, the {\Lyd} was
not captured (fell between segments), and the higher order Lyman
series lines were captured on Segment B.

\section{Image Analysis: Galaxy Properties}
\label{sec:imanal}

A re-analysis\footnote{We note that an incorrect $k$-correction
resulted in an overestimate of $M_B$ and $L_B/L_B^{\ast}$ in
Paper~I\nocite{paper1}. We also converted all magnitudes to the AB
system.} of the {\it HST}/WFPC2 F702W image was undertaken, presented,
and fully described by \citet{cwc-anatomy}.  We adopt the measured
quantities from that work.

The G1 quasar-galaxy impact parameter is $D = 58$ kpc.  The galaxy
photometric properties are $M_B =-21.6$, $M_K = -23.0$, and $B-K =
1.4$ (AB).  The $B$-band luminosity is
$L_B/L_B^{\ast} = 1.28$, where we use $M_B^{\ast}$ from the fit with
redshift reported by \citet{faber07}.  From an analysis using GIM2D
\citep{simard02}, we measure a half-light radius of $r_h = 4.2$ kpc,
disk scale length of $r_d = 0.9$ kpc, and bulge-to-total ratio of $B/T
= 0.99$.  The galaxy classifies as an E/S0 based upon its C-A
morphology \citep{abraham96}.  The galaxy inclination is $i =
15.9^{\circ}$, and the angle between the quasar sight line and the
major-axis of the projected ellipse of the galaxy is $\theta =
22.1^{\circ}$. For the properties of galaxy G2, see
\citet{cwc-anatomy} and our companion paper \citep{ggk-q1317}.

Photometric analysis of the ground-based images yielded dust-, color-,
and seeing-corrected AB apparent magnitudes for galaxy G1 of $m_{g'} =
23.4 \pm 0.03$, $m_{r'} = 22.2 \pm 0.03$ $m_{i'} = 21.01 \pm 0.03$ and
$m_{K_s} = 19.4 \pm 0.1$.  
A color composite image of the ground-based images can be viewed in
our companion paper \citep{ggk-q1317}, from which it can be
ascertained that galaxy G1 is clearly redder than galaxy G2 and that
the other galaxies in field (see
Figures~\ref{fig:data}$a$-\ref{fig:data}$d$) are likely at
substantially different redshifts.

For galaxy G3, we measured an F702W apparent magnitude of $m = 23.4
\pm 0.2$.  This value is based upon a dithered co-added image
constructed by A. Shapley, in which she removed the quasar via point
spread function (PSF) subtraction.  The quoted uncertainty is
statistical based upon the sky background; due to the PSF subtraction,
the error could be substantially underestimated.  Assuming galaxy G3
is at the redshift of the {\HI} absorption complex, i.e., $z\simeq
0.672$, and adopting the quoted apparent magnitude, we compute $M_B =
-19.3$ ($L_B/L_B^{\ast} = 0.16$) and $M_r=-20.7$ (SDSS $r$-band).  At
this redshift, the impact parameter to G3 is $D = 14.6$ kpc.

\begin{figure}[tbh]
\epsscale{1.2}
\plotone{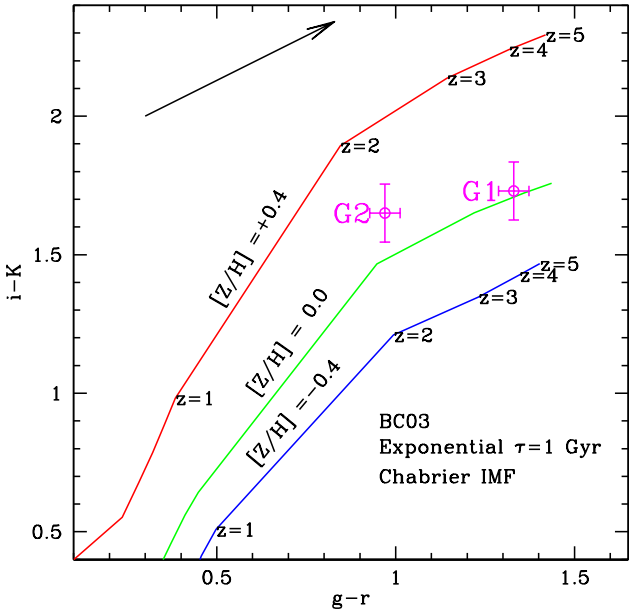}
\caption{The $g'\!-\!r'$ versus $i'\!-\!K_s$ color-color diagram
showing the measured dust-corrected colors of galaxy G1 and G2.  The
curves are \citet{bruzual03} stellar population models for
metallicities $[\hbox{Z/H}] = -0.4$, $0.0$, and $+0.4$ assuming an
exponential star formation rate with an $e$-folding time of 1 Gyr, and
a \citet{chabrier03} initial mass function.  The arrow provides the
reddening vector for $E(B\!-\!V) = 0.1$ in the rest frame of the
galaxies.  
}
\label{fig:color-color}
\end{figure}

In Figure~\ref{fig:color-color}, we plot the $g'\!-r\!'$ colors versus
$i'\!-\!K_s$ colors for G1 and G2.
Following \citet{bell03}, \citet{fontana04}, and \citet{swindle11}, we
used stellar population models to determine stellar masses,
$M_{\ast}$, of galaxies G1 and G2 from the observed colors.  We
employed the stellar population models of \citet{bruzual03} assuming a
\citet{chabrier03} initial mass function and an exponential star
formation history with an $e$-folding time of 1 Gyr\footnote{The
stellar population models were generated using the web service EZGAL
({\it www.baryons.org/ezgal\/}).}.  Also shown in
Figure~\ref{fig:color-color} are the locust of observed colors as a
function redshift for the \citet{bruzual03} stellar population models
for the metallicities $[\hbox{Z/H}] = -0.4$, $0.0$, and $+0.4$.  We
find galaxy G1 is consistent with a $\simeq 5.8$ Gyr old, solar
metallicity stellar population with a formation epoch of $z=4$.  The
stellar population models also yield the galaxy $K$-band mass to light
ratio, $(M/L)_{\hbox{\tiny $K$}}$.  From this ratio and the $K$-band
magnitude, we estimate $\log M_{\ast}/M_{\odot} = \log
(M/L)_{\hbox{\tiny $K$}} - 0.4 (M_K - M_{K,\odot}) = 11.5$ for galaxy
G1, where $M_{K,\odot} = -3.28$ is the solar value.


Using the technique of halo abundance matching, the galaxy virial
mass, $M_{\rm vir}$, can be estimated from the stellar mass
\citep[e.g.,][]{conroy09,behroozi10,moster10,stewart11}.  Abundance
matching assumes a monotonic functional relation between $M_{\rm vir}$
and $M_{\ast}$ by assigning the number of halos with $M_{\rm halo} >
M_{\rm vir}$ equal to the number of galaxies with $M_{\rm gal} >
M_{\ast}$.  As such, it matches the halo mass and stellar mass
functions globally with a roughly 0.25 dex uncertainty in $M_{\rm
vir}$ at fixed $M_{\ast}$, primarily due to the systematics in
estimates of $M_{\ast}$ \citep*{behroozi10}.  We employed the
parameterized functions presented by \citep{stewart11}.

\begin{deluxetable}{lccc}
\tablecolumns{4}
\tablewidth{0pt}
\tablecaption{Galaxy Properties \label{tab:galprops}}
\tablehead{
\colhead{Property}            & 
\colhead{G1}                  & 
\colhead{G2\tablenotemark{a}} & 
\colhead{G3\tablenotemark{b}} }
\startdata
$M_{\rm \ast}/M_{\odot}$ & $3\times 10^{11}$   & $1\times 10^{11}$   & $1\times 10^{10}$  \\
$M_{\rm vir}/M_{\odot}$  & $5\times 10^{13}$   & $8\times 10^{12}$   & $8\times 10^{11}$  \\
$T_{\rm vir}$, K          & $1\times 10^{7}$    & $3\times 10^{6} $   & $7\times 10^{5}$   \\
$v_{\rm circ}$, {\kms}    & 550                 & 280                 & 150                \\
$R_{\rm vir}$, kpc        & 750                 & 380                 & 180                \\
$M_{\rm gas}/M_{\odot}$  & $2\times 10^{10}$   & $1\times 10^{10}$   & $7\times 10^{9}$   \\
$M_{\rm bary}/M_{\odot}$ & $3.2\times 10^{11}$ & $1.1\times 10^{11}$ & $1.7\times 10^{10}$  \\
$f_g$                    & 0.05                & 0.09                & 0.34               
\enddata
\tablenotetext{a}{Taken from \citet{ggk-q1317}.}
\tablenotetext{b}{Assuming $z=0.672$.}
\end{deluxetable}

For galaxy G1, we obtained $M_{\rm vir} =
10^{13.7\!-\!14.4}$~M$_{\odot}$, where the lower value is given by the
\citet{conroy09} and \citet{moster10} fits and the higher value is
from the \citet*{behroozi10} fit\footnote{Abundance matching is the
most accurate for $M_{\rm vir} = 10^{11\!-\!13}$~M$_{\odot}$.  For less
massive halos, the stellar mass function is not tightly constrained.
For higher mass halos, particularly massive ellipticals, there is
substantial scatter between the published abundance matching
predictions/relations.  Thus, for a single case, abundance matching is
not highly robust for mapping $M_{\rm vir}$ from $M_{\ast}$ when $
M_{\rm vir} > 10^{13}$~M$_{\odot}$. \citep[][private
communication]{stewart12}}.  We adopted the \citet{conroy09} and
\citet{moster10} values $M_{\rm vir} = 10^{13.7}$~M$_{\odot}$.

Under the assumption that galaxy G3 resides at $z=0.672$, we use the
$\Lambda$CDM Bolshoi Simulation Database of \citet{trujillo11} to
estimate a virial mass of $M_{\rm vir} = 10^{11.9}$~M$_{\odot}$ using
abundance matching.  This mass is the average of 18,500 galaxies in
the absolute magnitude bin $-21.7 \leq M_r \leq -19.7$ with average
$M_r = -20.7$.  The abundance matching in this database is constrained
by the observed luminosity-circular velocity relation, the baryonic
Tully-Fisher relation, and the circular velocity function, allowing
all types of galaxies to be included.  Using the parameterized
abundance matching function for $M_{\ast}(M_{\rm vir},z)$ of
\citep{moster10}, we estimate that G3 has a stellar mass of $M_{\ast}
\simeq 10^{10.2}$~M$_{\odot}$.

The average gas mass, $M_{\rm gas}$, of a galaxy with stellar mass
$M_{\ast}$ can be estimated using the parameterized relation of
\citet{stewart11} based upon the baryonic Tully-Fisher relation study
of \citet{mcgaugh05}, the stellar, gas, and dynamical mass relation of
\citet{erb06}, and galaxy gas fraction stellar mass study of
\citet{stewart09}.  We obtain $M_{\rm gas} = 10^{10.2\!\pm
\!0.2}$~M$_{\odot}$ for galaxy G1\footnote{Estimates of the gas mass is
based upon the $M_{\rm gas}$-$M_{\ast}$ correlation for disk galaxies,
and is therefore not directly applicable to elliptical galaxies.
However, as shown in Fig.~2 of \citet{stewart09}, for $M_{\ast} >
10^{11}$~M$_{\odot}$, the gas-poor disk galaxy gas fractions reasonably
match those of the red galaxy sample of \citet{kannappan04}.  Thus, we
can crudely apply the relation to $M_{\ast} > 10^{11}$~M$_{\odot}$
elliptical galaxies \citep[][private communication]{stewart12}.} and
estimate $M_{\rm gas} = 10^{9.9\!\pm \!0.3}$ ~M$_{\odot}$ for G3.
Thus, we deduce averaged baryonic gas fractions, $f_g = M_{\rm
gas}/(M_{\ast}+M_{\rm gas})$, of 5\% for G1 and 34\% for G3.

The virial radii, virial temperatures, and circular velocities for
galaxies G1 and G3 were computed using the relations of
\citet{bryan98} for $\Omega _{\hbox{\tiny R}} = 0$.  For galaxy G1, we
obtained $R_{\rm vir} = 750$ kpc, $T_{\rm vir} = 1 \times 10^7$ K, and
$v_{\rm circ} = 550$ {\kms} and for galaxy G3, we obtained $R_{\rm
vir} = 180$ kpc, $T_{\rm vir} = 7 \times 10^5$ K, and $v_{\rm circ} =
150$ {\kms}.  In Table~\ref{tab:galprops}, we summarize the deduced
galaxy properties, which are representative averages for galaxies with
their observed photometric properties.

\section{Spectral Analysis: Absorption Properties}
\label{sec:specanal}

In the FOS spectrum, a very broad {\HI} absorption complex was
observed \citep[first reported by][]{cat1}.  In Paper~I\nocite{paper1}
we fitted the {\Lya} with five Gaussian components at redshifts
$z=0.66914$, 0.67157, 0.67355, 0.67559, and 0.67707, respectively.
The total rest-frame equivalent width was determined to be
$W_r({\Lya}) = 2.87$~{\AA} over a rest-frame velocity spread of
1420~{\kms}.  From the STIS spectrum, the $3~\sigma$ {\CIV}~$\lambda
1548$ equivalent width limit was estimated to be $W_r(1548) \leq
0.03$~{\AA} at $z = 0.6719$.  In the FOS spectrum, the $3~\sigma$
{\OVI}~$\lambda 1031$ equivalent width limit was estimated to be
$W_r(1031) \leq 0.21$~{\AA}.  For {\MgII}~$\lambda 2796$ in the HIRES
spectrum, we obtained $W_r(2796) \leq 7$~m{\AA} to $3~\sigma$.  These
values assumed unresolved absorption.

For this work, we objectively locate absorption lines or place limits
on their equivalent widths employing the optimized methods of
\citet{schneider93} and \citet{weakI} as modified using the methods
for unresolved lines and pattern noise developed by \citet{lawton08}.
We adopt a $5~\sigma$ detection threshold and quote $3~\sigma$ limits.

\subsection{Neutral Hydrogen Lines}

\begin{figure}[thb]
\epsscale{1.25}
\plotone{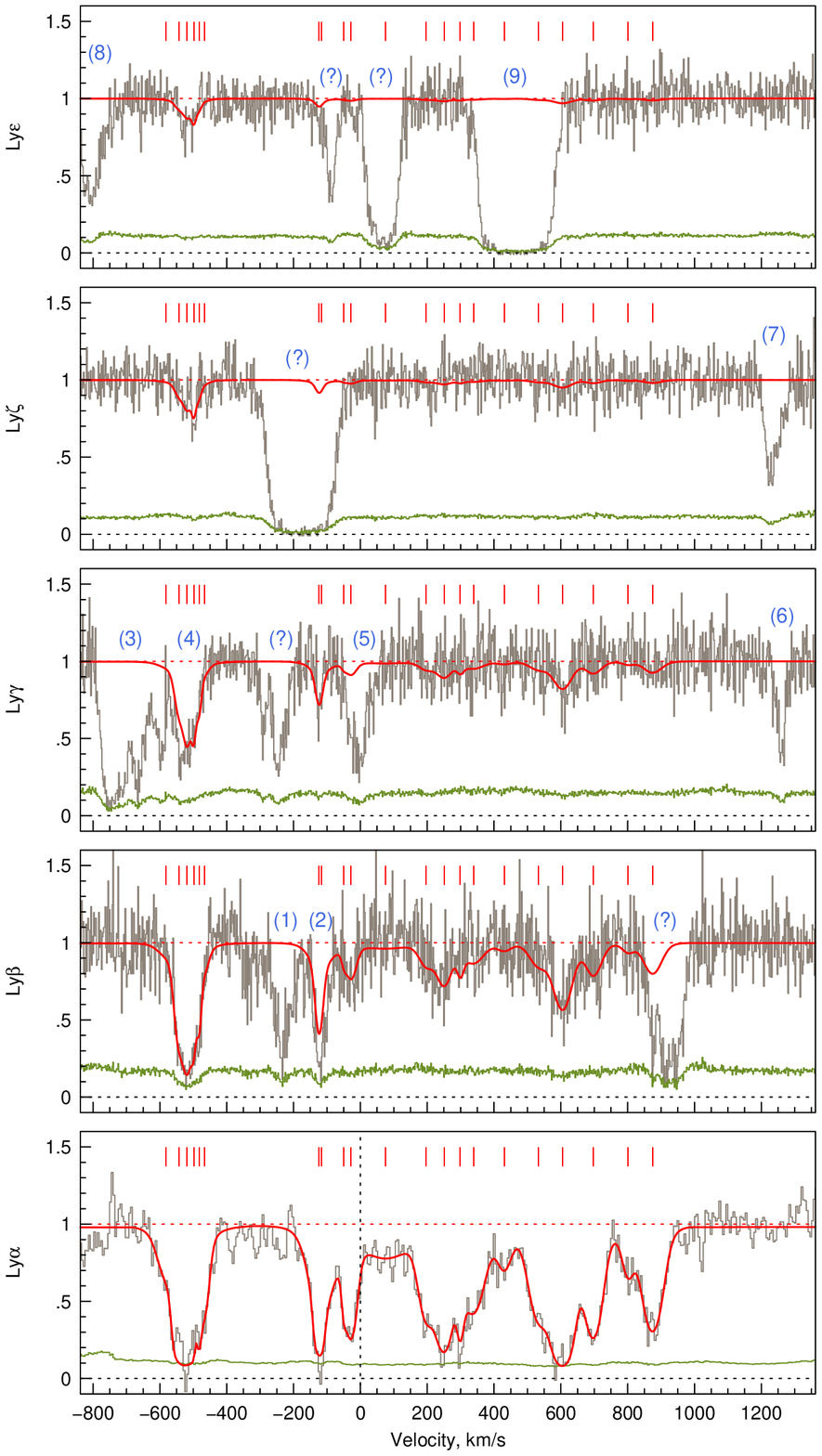}
\caption{Voigt profile decomposition of the {\HI} complex,
including {\Lyb}, {\Lyg}, {\Lyz}, and {\Lye} ({\Lyd} was not
captured).  The vertical dashed line is the redshift of galaxy G1.
The component parameters centroids are marked by ticks above the
continuum and the fitted component parameters are listed in
Tables~\ref{tab:hiresult} and \ref{tab:metals}.  Several blends are
present and identified as: (1,2) {\OVI} $\lambda 1031$ associated with
the $z=0.6610$ metal system; (3,4) {\CIII} $\lambda 977$ at
$z=0.6610$; (5) {\Lyd} at $z=0.7738$; (6) {\CIII} $\lambda 977$ from
clouds \#7 \& \#8 of the {\HI} complex; (7) {\Lyb} $z=0.5349$; (8)
Galactic {\CIV} $\lambda 1550$; (9) {\Lyz} at $z=0.6610$.  Those
marked with ``(?)'' may well be {\Lya} lines in that there is no
corroborating data to suggest they are metal lines nor higher order
Lyman series lines.}
\label{fig:HIfits}
\end{figure}

In Figure~\ref{fig:HIfits}, we present the {\Lya}, {\Lyb}, {\Lyg},
{\Lye}, and {\Lyz} absorption observed in the COS spectrum as a
function of rest-frame velocity relative to the $z_{\rm gal} = 0.6719$
galaxy.  No {\HI} absorption was detected for Lyman series lines
higher than {\Lyz}.  There is significant blending in the {\Lyb},
{\Lyg}, {\Lye}, and {\Lyz} lines, which are identified in
Figure~\ref{fig:HIfits}, when possible.

The curves through the data are $\chi^2$ minimized Voigt profile (VP)
fits obtained using our code MINFIT \citep{cwc-thesis,cv01,cvc03}.
The ticks above the normalized continuum provide the VP component
velocities.  During the fitting, the COS instrumental line spread
function (ISF) was convolved with the VP model.  The COS ISF
appropriate for the spectrograph settings and the observed wavelength
of each transition was determined via interpolation of the on-line
tabulated data \citep[cf.,][]{cos-ihb,kriss11}.


For the VP fitting, pixels compromised by blending were masked out of
the least-squares fit vector.  We present the results of our VP
modeling in Table~\ref{tab:hiresult}.  We fitted a total of 21
components, or ``clouds'' (MINFIT returns the minimum number of
components based upon their statistical significance through a series
of F-tests and confidence level checks).  Assuming thermal broadening,
we converted the Doppler $b$ parameter into the ``cloud'' temperature.

\begin{deluxetable}{lcrccc}
\tablecolumns{6}
\tablewidth{0pt}
\tablecaption{{Voigt Profile Decomposition of the \HI} Complex\label{tab:hiresult}}
\tablehead{
\colhead{Cld\#}       & 
\colhead{$z_{cl}$}    & 
\colhead{$v_{cl}$\tablenotemark{a}}    & 
\colhead{$\log N({\HI})$}  & 
\colhead{$b({\HI})$}  & 
\colhead{$T$} \\
\colhead{\phantom{X}} & 
\colhead{\phantom{X}} & 
\colhead{[{\kms}]}     & 
\colhead{\phantom{X}} & 
\colhead{[{\kms}]}    & 
\colhead{[$10^3$ K] } }
\startdata
   1\tablenotemark{b}  & 0.668645 & $ -582.1$ & $13.24\pm0.59$ & $31.9\pm31.3$ & $62.0 \pm 60.9$ \\  
   2\tablenotemark{b}  & 0.668861 & $ -543.5$ & $14.26\pm0.54$ & $14.7\pm 9.9$ & $13.2 \pm  8.9$ \\  
   3\tablenotemark{b}  & 0.669000 & $ -518.6$ & $14.53\pm0.33$ & $12.8\pm10.7$ & $10.0 \pm  8.4$ \\    
   4\tablenotemark{b}  & 0.669112 & $ -498.5$ & $14.51\pm0.14$ & $ 3.2\pm 2.1$ & $ 0.6 \pm  0.4$ \\    
   5\tablenotemark{b}  & 0.669199 & $ -482.9$ & $14.07\pm0.20$ & $ 4.0\pm 1.8$ & $ 1.0 \pm  0.4$ \\    
   6\tablenotemark{b}  & 0.669290 & $ -520.6$ & $13.38\pm0.38$ & $15.5\pm11.2$ & $14.7 \pm 10.5$ \\      
   7                   & 0.671203 & $ -123.8$ & $14.13\pm0.10$ & $11.5\pm 2.9$ & $8.1  \pm  2.0$ \\
   8                   & 0.671244 & $ -116.6$ & $13.69\pm0.14$ & $43.3\pm14.4$ & $114.5\pm  38.1$ \\
   9                   & 0.671617 & $  -49.6$ & $12.93\pm0.94$ & $ 5.8\pm15.5$ & $2.1  \pm   5.5$ \\
  10                   & 0.671739 & $  -27.7$ & $13.70\pm0.20$ & $18.3\pm 7.7$ & $20.5 \pm   8.6$ \\
  11                   & 0.672319 & $  +76.1$ & $13.43\pm0.19$ & $88.6\pm49.2$ & $479.3\pm 266.1$ \\
  12                   & 0.672994 & $ +197.0$ & $13.59\pm0.28$ & $26.4\pm11.9$ & $42.5 \pm  19.2$ \\
  13                   & 0.673291 & $ +250.3$ & $13.93\pm0.14$ & $27.6\pm10.9$ & $46.4 \pm  18.3$ \\
  14                   & 0.673562 & $ +289.9$ & $13.38\pm0.26$ & $10.5\pm 6.9$ & $6.7  \pm   4.4$ \\
  15                   & 0.673787 & $ +339.1$ & $13.68\pm0.15$ & $38.4\pm14.1$ & $89.9 \pm  33.0$ \\
  16                   & 0.674297 & $ +430.5$ & $13.11\pm0.12$ & $26.1\pm10.1$ & $41.6 \pm  16.1$ \\
  17                   & 0.674872 & $ +533.5$ & $13.70\pm0.13$ & $34.6\pm 8.8$ & $73.0 \pm  18.6$ \\
  18                   & 0.675272 & $ +605.1$ & $14.26\pm0.05$ & $33.7\pm 4.2$ & $69.4 \pm   8.6$ \\
  19                   & 0.675789 & $ +698.8$ & $13.82\pm0.04$ & $29.0\pm 3.4$ & $51.5 \pm   5.9$ \\
  20                   & 0.676366 & $ +801.2$ & $13.10\pm0.12$ & $20.5\pm 7.9$ & $25.7 \pm   9.9$ \\
  21                   & 0.676776 & $ +874.7$ & $13.86\pm0.04$ & $33.5\pm 3.6$ & $68.3 \pm   7.3$ 
\enddata
\tablenotetext{a}{Velocities are measured with respect to $z_{\rm gal}=0.6719$.}
\tablenotetext{b}{See text for discussion of these components, which
 required a deblending treatment.  We also fitted this region
 with a single VP component at $v=-516.9$, with $\log
 N= 14.88\pm0.02$, and $b = 32.76\pm0.861$ {\kms}.  However,
 the VP model provided a poor match the structure in the core
 of the {\Lya} feature.}
 \end{deluxetable}

\begin{figure}[thb]
\epsscale{1.20}
\plotone{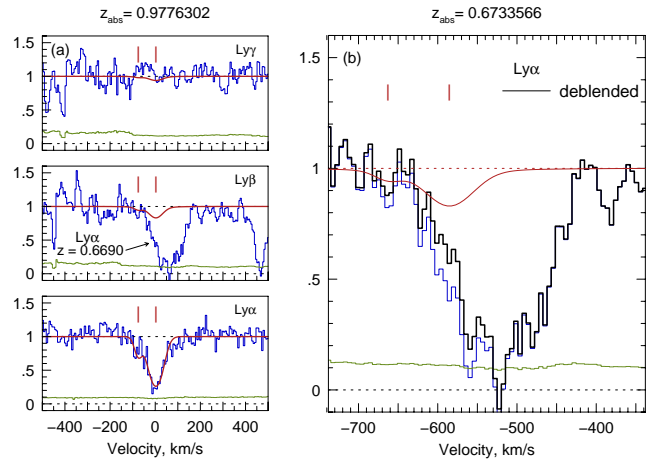}
\caption{($a$) The {\Lyb} line (COS) from the $z_{abs}=0.9778$ {\Lya}
absorber (center panel) is blended with the $v \simeq -530$~{\kms}
absorption of the {\HI} complex.  The solid curves are a
two-component Voigt profile model of the {\HI} absorption centered
at $z_{abs}=0.97763$, including the constraint from {\Lya} (STIS)
and {\Lyg} (COS).  --- ($b$) Subtraction of the $z_{abs}=0.9776$
{\Lyb} Voigt profile model from the affected component of the 
{\HI} absorption complex.}
\label{fig:deblend}
\end{figure}

The strong {\Lya} absorption centered at $v \simeq -530$~{\kms} ($z
\simeq 0.6690$ ) is blended with {\Lyb} at $z=0.9776$ associated with
the {\Lya} absorption identified by \citet{cat2} at $z=0.9778$ in the
FOS spectrum, which we confirmed in the STIS spectrum.  Prior to VP
fitting the {\HI} complex, we deblended the $v \simeq -530$~{\kms}
{\Lya} feature using VP modeling and employing both the COS and STIS
ISFs.  In Figure~\ref{fig:deblend}$a$, we present the VP fits at
$z=0.97763$.  The flux decrement of the {\Lyb} VP model was then
subtracted from that of the $v \simeq -530$~{\kms} {\Lya} absorption
from the {\HI} complex.  The resulting deblended profile is
illustrated in Figure~\ref{fig:deblend}$b$.

Prior to and following the deblending process, we experienced
difficulty obtaining a satisfactory simultaneous fit to the {\Lya}
and {\Lyb} absorption in the $v \simeq -530$~{\kms} feature.
Though the deblending provided an improved VP model, we caution that
the number of VP components and the resulting column densities and
Doppler $b$ parameters we adopted should be viewed with discretion
(see comments for Table~\ref{tab:hiresult}).

\subsection{Metal Lines}

\begin{figure}[thb]
\plotone{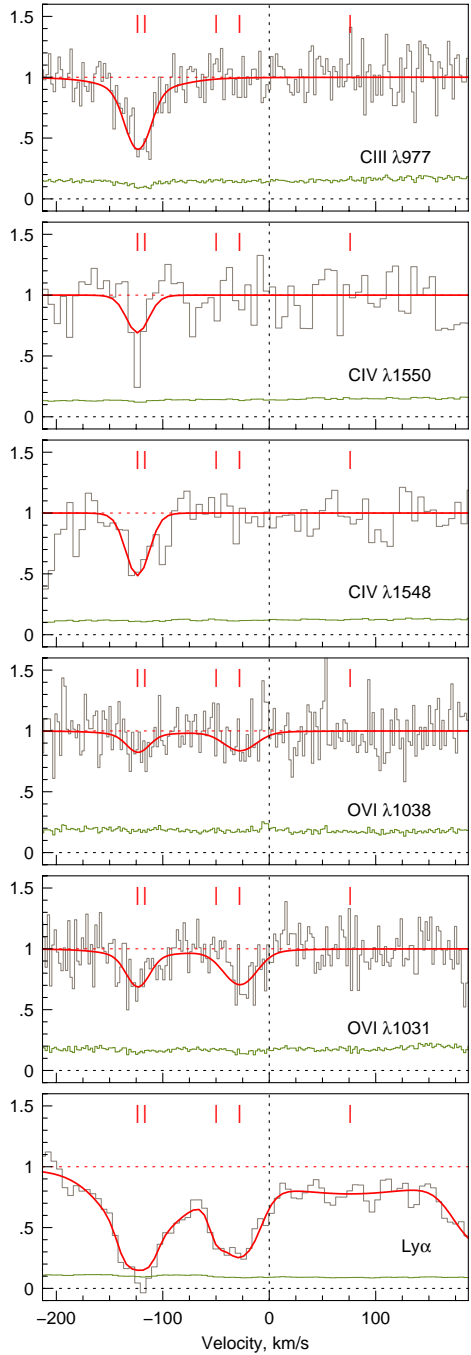}
\caption{Expanded region of the Voigt profile decomposition of the
detected metal lines over the redshift window for clouds \#7 through
\#11 of the {\HI} complex. The vertical dashed line is the
redshift of galaxy G1.}
\label{fig:metals}
\end{figure}

We examined the COS, STIS, and HIRES quasar spectra for associated
metal-line transitions, including the {\MgIIdblt}, {\SiIVdblt},
{\CIVdblt}, and {\OVIdblt} zero-volt resonance doublets, and transitions
from {\SiII}, {\SiIII}, {\CII}, {\CIII}, etc.  In
Figure~\ref{fig:metals}, we present the wavelength regions of the
quasar spectra corresponding to detected metal lines {\CIII} $\lambda
977$ (COS), {\CIVdblt} (STIS), and {\OVIdblt} (COS).  We also show the
{\Lya} absorption over the same velocity range. VP components \#7
through \#11 are shown, where the ``cloud'' number is given in column
1 of Table~\ref{tab:hiresult}.  Only clouds \#7, \#8, and \#10 have
detected metals.

In Paper~I\nocite{paper1}, we reported no detected metal absorption in
{\CIV} (STIS) and {\OVI} (FOS).  However, we clearly detected {\OVI}
absorption in the COS spectrum in the velocity range $-200 \leq v \leq
0$~{\kms} relative to $z_{\rm gal} = 0.6719$.  We also detected
{\CIII}~$\lambda 977$ absorption at $v \simeq - 120$~{\kms}.  We then
re-examined the {\CIV} absorption, and formally detected
{\CIV}~$\lambda 1548$ aligned in velocity with the {\OVI} and {\CIII}
absorption.  The {\CIV}~$\lambda 1550$ absorption, which we had
interpreted as noise in Paper~I\nocite{paper1}, is detected at the
$3.5~\sigma$ level.

In Figure~\ref{fig:metals}, the curves through the data are fitted VP
components.  For these fits, we modified MINFIT to hold the VP
component redshifts (velocities) and Doppler $b$ parameters constant,
allowing only the column densities to be $\chi^2$ minimized, and
fitted the metal lines using the velocities and Doppler $b$ parameters
from the VP fits to the {\HI} Lyman series lines.  As with the {\HI}
fits described above, the COS ISF \citep{cos-ihb,kriss11} and the STIS
ISF \citep{stis-ihb} appropriate for the spectrograph settings and the
observed wavelength of each transition was determined via
interpolation of the on-line tabulated data.


When a VP component column density was objectively determined to be
insignificant, MINFIT returned an upper limit.  Several tests for
significance were conducted during the least-square fitting
convergence; in short, for a column density to be deemed significant,
the fitted VP value and its uncertainty had to be inconsistent with
the upper limit on the column density (in that precise region of the
spectrum and for that Doppler $b$ parameter), and the integrated
apparent optical depth \citep{savage91} had to be consistent with the
VP component and inconsistent with the the upper limit on the column
density.

In Table~\ref{tab:metals}, we list the metal-line absorption
properties for all clouds.  In MINFIT, the equivalent width limits are
determined directly from the spectra using the methods of
\citet{schneider93} and \citet{weakI}, but modified for partially or
fully resolved features \citep[see][]{lawton08}.  For each transition,
we use the Doppler $b$ parameter from the VP fit to the {\HI} series
convolved with the appropriate ISF to measure the upper limit on the
equivalent width in each cloud.  The uncertainty in the Doppler $b$
width is used to determine the $1~\sigma$ spread in this limit.  The
column density limits and the $1~\sigma $ spread in these are
determined from the curve of growth.

\section{Ionization Modeling}
\label{sec:models}

We employed 
our own photo+collisional ionization code \citep{cwc-rates}, which is
very similar to the code LINESPEC \citep{verner90}.  The code treats
photoionization, Auger ionization, direct collisional ionization,
excitation-autoionization, photo-recombination, high and low
temperature dielectronic recombination, charge transfer ionization by
H$^{+}$, and charge transfer recombination by H$^0$ and He$^0$.
Metals up to zinc can be incorporated, and all ionization stages for
each elemental species are modeled.  The code is appropriate for
optically thin gas in which no ionization structure is present.

\begin{deluxetable*}{lcrrrrrrr}
\tablecolumns{9}
\tablewidth{0pc}
\tablecaption{Selected Metal-Line Measurements \label{tab:metals}}
\tablehead{
\colhead{Cld\#}              & 
\colhead{$z_{cl}$}           & 
\colhead{$v_{cl}$\tablenotemark{a}}    & 
\colhead{$\log N({\MgII})$}  & 
\colhead{$\log N({\SiIII})$} & 
\colhead{$\log N({\SiIV})$}  & 
\colhead{$\log N({\CIII})$}  & 
\colhead{$\log N({\CIV})$}   & 
\colhead{$\log N({\OVI})$}   \\
\colhead{\phantom{X}} & 
\colhead{\phantom{X}} & 
\colhead{[{\kms}]}     & 
\colhead{\phantom{X}} & 
\colhead{\phantom{X}} & 
\colhead{\phantom{X}} & 
\colhead{\phantom{X}} & 
\colhead{\phantom{X}} & 
\colhead{\phantom{X}} }
\startdata
  1& 0.668645 & $ -582.1$
   &  $<\!\!11.70_{-0.30}^{+0.17}$ 
   &  $<\!\!12.46_{-9.99}^{+0.16}$ 
   &  $<\!\!13.07_{-9.99}^{+0.15}$ 
   &  $<\!\!12.77_{-9.99}^{+0.16}$ 
   &  $<\!\!13.26_{-9.99}^{+0.16}$ 
   &  $<\!\!13.58_{-9.99}^{+0.16}$ \\[2pt]
  2& 0.668861 & $ -543.5$
   &  $<\!\!11.53_{-0.18}^{+0.13}$ 
   &  $<\!\!12.29_{-0.16}^{+0.11}$ 
   &  $<\!\!12.92_{-0.13}^{+0.10}$ 
   &  $<\!\!12.60_{-0.15}^{+0.11}$ 
   &  $<\!\!13.12_{-0.16}^{+0.11}$ 
   &  $<\!\!13.42_{-0.14}^{+0.11}$ \\[2pt]
  3& 0.669000 & $ -518.6$
   &  $<\!\!11.50_{-0.25}^{+0.15}$ 
   &  $<\!\!12.26_{-0.15}^{+0.13}$ 
   &  $<\!\!12.90_{-9.99}^{+0.12}$ 
   &  $<\!\!12.57_{-0.06}^{+0.14}$ 
   &  $<\!\!13.09_{-0.12}^{+0.13}$ 
   &  $<\!\!13.40_{-0.03}^{+0.13}$ \\[2pt]
  4& 0.669112 & $ -498.5$
   &  $<\!\!11.21_{-0.07}^{+0.12}$ 
   &  $<\!\!12.02_{-9.99}^{+0.08}$ 
   &  $<\!\!12.79_{-9.99}^{+0.01}$ 
   &  $<\!\!12.41_{-9.99}^{+0.05}$ 
   &  $<\!\!12.87_{-9.99}^{+0.08}$ 
   &  $<\!\!13.24_{-9.99}^{+0.01}$ \\[2pt]
  5& 0.669199 & $ -482.9$
   &  $<\!\!11.25_{-0.11}^{+0.09}$ 
   &  $<\!\!12.05_{-0.06}^{+0.06}$ 
   &  $<\!\!12.78_{-9.99}^{+0.03}$ 
   &  $<\!\!12.39_{-9.99}^{+0.04}$ 
   &  $<\!\!12.90_{-9.99}^{+0.06}$ 
   &  $<\!\!13.22_{-9.99}^{+0.04}$ \\[2pt]
  6& 0.669290 & $ -520.6$
   &  $<\!\!11.54_{-0.20}^{+0.13}$ 
   &  $<\!\!12.30_{-0.17}^{+0.12}$ 
   &  $<\!\!12.93_{-0.14}^{+0.11}$ 
   &  $<\!\!12.61_{-0.16}^{+0.12}$ 
   &  $<\!\!13.13_{-0.17}^{+0.12}$ 
   &  $<\!\!13.43_{-0.16}^{+0.12}$ \\[2pt]
  7& 0.671203 & $ -123.8$
   &  $<\!\! 11.48_{-0.06}^{+0.05}$ 
   &  $<\!\!12.24_{-0.05}^{+0.04}$ 
   &  $<\!\!12.89_{-0.04}^{+0.04}$ 
   &  $ 13.24\pm 0.07$                          
   &  $ 13.40\pm 0.08$                          
   &  $ 13.49\pm 0.13$ \\[2pt]                  
  8& 0.671244 & $ -116.6$
   &  $<\!\!11.76_{-0.08}^{+0.07}$ 
   &  $<\!\!12.53_{-0.07}^{+0.06}$ 
   &  $<\!\!13.12_{-0.07}^{+0.06}$ 
   &  $12.42\pm 0.39$                           
   &  $<\!\!13.33_{-0.07}^{+0.06}$ 
   &  $ 12.89\pm 0.68$ \\[2pt] 
  9& 0.671617 & $  -49.6$
   &  $<\!\!11.33_{-0.18}^{+0.38}$ 
   &  $<\!\!12.10_{-0.17}^{+0.31}$ 
   &  $<\!\!12.81_{-0.16}^{+0.24}$ 
   &  $<\!\!12.42_{-0.09}^{+0.31}$ 
   &  $<\!\!12.95_{-0.15}^{+0.31}$ 
   &  $<\!\!13.25_{-0.11}^{+0.30}$ \\[2pt]
 10& 0.671739 & $  -27.7$
   &  $<\!\!11.58_{-0.10}^{+0.08}$ 
   &  $<\!\!12.33_{-0.09}^{+0.07}$ 
   &  $<\!\!12.96_{-0.08}^{+0.07}$ 
   &  $<\!\!12.65_{-0.09}^{+0.07}$ 
   &  $<\!\!13.16_{-0.09}^{+0.07}$ 
   &  $ 13.62\pm 0.08$ \\[2pt] 
 11& 0.672319 & $  +76.1$
   &  $<\!\!11.91_{-0.14}^{+0.11}$ 
   &  $<\!\!12.72_{-0.13}^{+0.10}$ 
   &  $<\!\!13.25_{-0.13}^{+0.10}$ 
   &  $<\!\!12.98_{-0.14}^{+0.10}$ 
   &  $<\!\!13.47_{-0.14}^{+0.10}$ 
   &  $<\!\!13.78_{-0.14}^{+0.10}$ \\[2pt]
 12& 0.672994 & $ +197.0$
   &  $<\!\!11.65_{-0.11}^{+0.09}$ 
   &  $<\!\!12.41_{-0.10}^{+0.08}$ 
   &  $<\!\!13.03_{-0.09}^{+0.07}$ 
   &  $<\!\!12.73_{-0.10}^{+0.08}$ 
   &  $<\!\!13.23_{-0.10}^{+0.08}$ 
   &  $<\!\!13.54_{-0.10}^{+0.08}$ \\[2pt]
 13& 0.673291 & $ +250.3$
   &  $<\!\!11.66_{-0.10}^{+0.08}$ 
   &  $<\!\!12.42_{-0.09}^{+0.07}$ 
   &  $<\!\!13.04_{-0.08}^{+0.07}$ 
   &  $<\!\!12.74_{-0.09}^{+0.07}$ 
   &  $<\!\!13.24_{-0.09}^{+0.07}$ 
   &  $<\!\!13.55_{-0.09}^{+0.07}$ \\[2pt]
 14& 0.673562 & $ +289.9$
   &  $<\!\!11.46_{-0.18}^{+0.12}$ 
   &  $<\!\!12.22_{-0.14}^{+0.10}$ 
   &  $<\!\!12.87_{-0.10}^{+0.09}$ 
   &  $<\!\!12.52_{-0.12}^{+0.11}$ 
   &  $<\!\!13.05_{-0.14}^{+0.10}$ 
   &  $<\!\!13.36_{-0.12}^{+0.11}$ \\[2pt]
 15& 0.673787 & $ +339.1$
   &  $<\!\!11.73_{-0.09}^{+0.07}$ 
   &  $<\!\!12.50_{-0.08}^{+0.07}$ 
   &  $<\!\!13.09_{-0.08}^{+0.06}$ 
   &  $<\!\!12.81_{-0.08}^{+0.07}$ 
   &  $<\!\!13.30_{-0.08}^{+0.07}$ 
   &  $<\!\!13.63_{-0.08}^{+0.07}$ \\[2pt]
 16& 0.674297 & $ +430.5$
   &  $<\!\!11.65_{-0.09}^{+0.08}$ 
   &  $<\!\!12.41_{-0.08}^{+0.07}$ 
   &  $<\!\!13.03_{-0.08}^{+0.06}$ 
   &  $<\!\!12.73_{-0.09}^{+0.07}$ 
   &  $<\!\!13.22_{-0.08}^{+0.07}$ 
   &  $<\!\!13.54_{-0.08}^{+0.07}$ \\[2pt]
 17& 0.674872 & $ +533.5$
   &  $<\!\!11.71_{-0.06}^{+0.05}$ 
   &  $<\!\!12.47_{-0.05}^{+0.05}$ 
   &  $<\!\!13.08_{-0.05}^{+0.04}$ 
   &  $<\!\!12.79_{-0.06}^{+0.05}$ 
   &  $<\!\!13.28_{-0.05}^{+0.05}$ 
   &  $<\!\!13.61_{-0.05}^{+0.05}$ \\[2pt]
 18& 0.675272 & $ +605.1$
   &  $<\!\!11.71_{-0.03}^{+0.03}$ 
   &  $<\!\!12.47_{-0.03}^{+0.02}$ 
   &  $<\!\!13.07_{-0.02}^{+0.02}$ 
   &  $<\!\!12.78_{-0.03}^{+0.02}$ 
   &  $<\!\!13.28_{-0.03}^{+0.02}$ 
   &  $<\!\!13.60_{-0.03}^{+0.02}$ \\[2pt]
 19& 0.675789 & $ +698.8$
   &  $<\!\!11.67_{-0.03}^{+0.02}$ 
   &  $<\!\!12.43_{-0.02}^{+0.02}$ 
   &  $<\!\!13.05_{-0.02}^{+0.02}$ 
   &  $<\!\!12.75_{-0.02}^{+0.02}$ 
   &  $<\!\!13.25_{-0.02}^{+0.02}$ 
   &  $<\!\!13.56_{-0.02}^{+0.02}$ \\[2pt]
 20& 0.676366 & $ +801.2$
   &  $<\!\!11.60_{-0.09}^{+0.08}$ 
   &  $<\!\!12.35_{-0.08}^{+0.07}$ 
   &  $<\!\!12.99_{-0.07}^{+0.06}$ 
   &  $<\!\!12.67_{-0.09}^{+0.07}$ 
   &  $<\!\!13.18_{-0.08}^{+0.07}$ 
   &  $<\!\!13.48_{-0.08}^{+0.07}$ \\[2pt]
 21& 0.676776 & $ +874.7$
   &  $<\!\!11.70_{-0.02}^{+0.02}$ 
   &  $<\!\!12.46_{-0.02}^{+0.02}$ 
   &  $<\!\!13.07_{-0.02}^{+0.02}$ 
   &  $<\!\!12.78_{-0.02}^{+0.02}$ 
   &  $<\!\!13.27_{-0.02}^{+0.02}$ 
   &  $<\!\!13.60_{-0.02}^{+0.02}$
\enddata
\tablenotetext{a}{Velocities are measured with respect to $z_{\rm gal}=0.6719$.}
\end{deluxetable*}


The input cloud parameters are the hydrogen number density, {\hden},
equilibrium kinetic temperature, $T$, and the mass fraction of metals,
$Z$.  Solar abundance mass fractions are taken from Table 1.4 of
\citet{draine11}, based upon \citet{asplund09}, though the code has
the option of modifying the relative abundances.  A \citet{haardt11}
ionizing spectrum is used for the ultraviolet background (UVB).
However, a stellar ionizing spectrum or a combined stellar plus UVB
ionizing spectrum can be incorporated \citep[see][for
details]{cwc-rates}.

The code obtains an initial guess solution for the density of each
ionic species based upon the assumption of adjacent ion stage
ionization and recombination balance (i.e., neglecting Auger and
charge transfer processes).  Then, the rate matrix is solved using the
code dqed.f, a Hanson/Krogh nonlinear least squares algorithm with
linear constraints based on quadratic-tensor local model \citep{dqed}.
The outputs of the ionization code are the electron density, and the
ionization and recombination rate coefficients, ionization fractions
and the number densities for all ionic species.

Since the galaxy G1 is 58 kpc from the location where the gas is
probed in absorption, and the stellar population is clearly dominated
by red stars, we assume a UVB-only ionizing spectrum.  We also assume
a solar abundance pattern and include metals up to iron (however, we
omitted lithium, beryllium, boron, fluorine, and the nobel gas
elements for which many of the ionization and recombination rates are
not well determined).

For a given cloud model, the resulting column density of ionic species
X is obtained by 
\begin{equation}
\log N({\rm X}) = \log N({\HI}) - \log f_{\hbox{\tiny HI}}  + 
\log n_{\hbox{\tiny X}} - \log \hbox{\hden} \, ,
\label{eq:colden}
\end{equation}
where $N({\HI})$ is the {\HI} column density from the VP fit to the
data, {\hden} is the input hydrogen density of the cloud model, and
where $f_{\hbox{\tiny HI}}$ is the ionization fraction of H$^{0}$ and
$n_{\hbox{\tiny X}}$ is the number density of ionic species X output
by the code, respectively.  Note that the metallicity of the model is
implied.  In practice, $n_{\hbox{\tiny X}}$ in a given model scales
directly in proportion to $Z$.  Thus, if the metallicity of the cloud
model is $Z_{\rm m}$, for which the resulting density of species X is
$n_{{\hbox{\tiny X}},{\rm m}}$, then $\log n_{\hbox{\tiny X}} = \log
n_{{\hbox{\tiny X}},{\rm m}} + (\log Z - \log Z_{\rm m})$.  For a
given {\hden}, such scaling of the models is valid only if
$f_{\hbox{\tiny HI}}$ is independent of metallicity, which holds for
$\log Z/Z_{\odot} \leq -1$.

Comparing Eq.~\ref{eq:colden} to the measured values of $N({\rm X})$
from our VP fitting, we can constrain the metallicity and {\hden} of
each cloud.  The range in these constrained quantities are based upon
the $1~\sigma$ uncertainties in the measured column densities.  Only
three clouds have detected metal lines and therefore measured $N({\rm
X})$.  These clouds are \#7, \#8, and \#10.  We examine these clouds
in the following subsections.

\subsection{Clouds \#7 and \#8}
\label{sec:clds78}

\begin{figure*}[thb]
\plotone{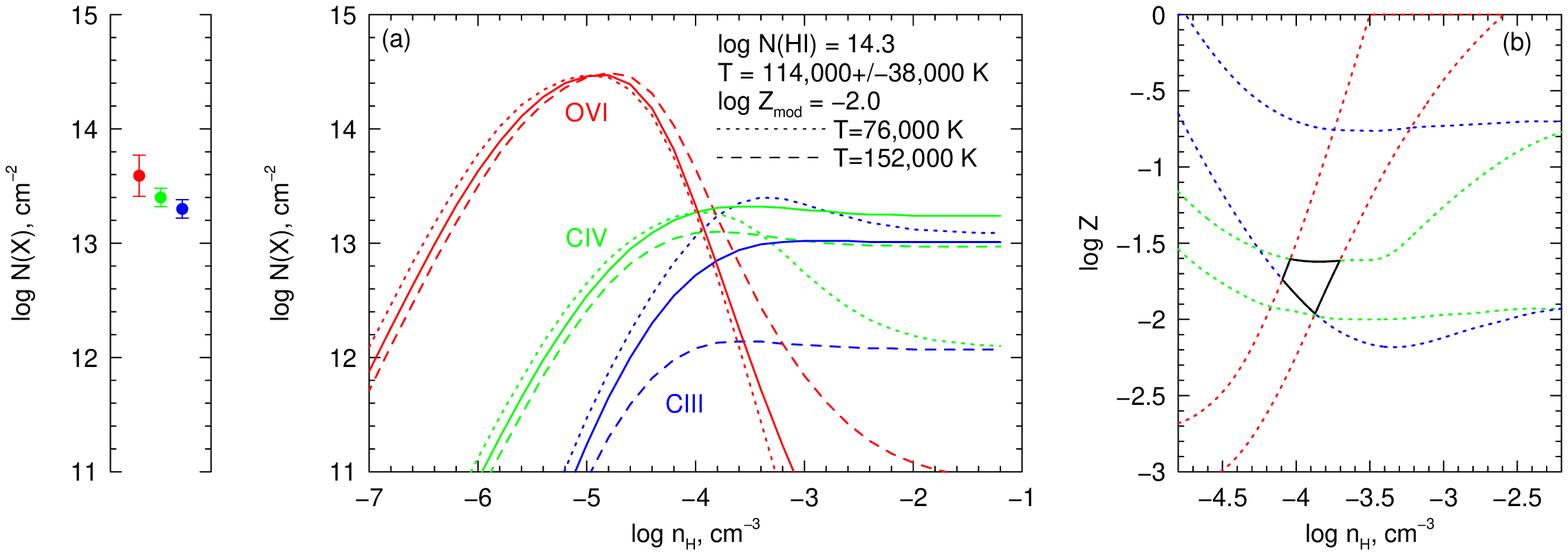}
\caption{($a$) Ionization models of clouds \#7+8 for a metal mass
fraction of $\log Z/Z_{\odot} = -2.0$ with solar abundance ratios and
temperature $T=114,000\pm 38,000$ K.  The solid curves are the model
column densities of {\OVI} (red), {\CIV} (green), and {\CIII} (blue)
and the dotted curves accounting for the uncertainties in the
temperature.  The measured values of $N({\OVI})$, $N({\CIV})$ and
$N({\CIII})$ are provided in the panel to the left. --- ($b$) The
$\log Z$--$\log n_{\hbox{\tiny H}}$ plane showing the constraints on
the models from the measured data (same color scheme).  Solid black
curves provide the overlap regions. }
\label{fig:cloud78}
\end{figure*}

\begin{figure*}[thb]
\plotone{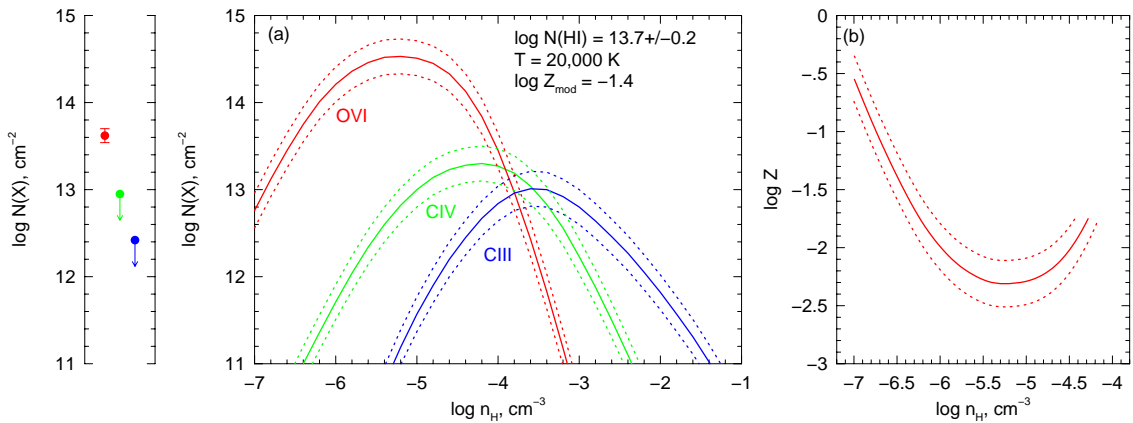}
\caption{($a$) Ionization model of cloud \#10 for a metal mass
fraction of $\log Z/Z_{\odot} = -1.4$ with solar abundance ratios and
temperature $T=20,000$ K.  The solid curves and data points are as
described in Figure~\ref{fig:cloud78}, whereas the dotted curves
account for a 0.2 dex error in $\log N({\HI})$.  --- ($b$) The
constrained range of the metal mass fraction on the $\log Z$--$\log
n_{\hbox{\tiny H}}$ plane, accounting for the column density limits on
the carbon ions.}
\label{fig:cloud10}
\end{figure*}


Clouds \#7 and \#8 are located at $v \simeq -120$~{\kms} with respect
to the systemic velocity of G1 and are separated by $\Delta v =
7$~{\kms}, which is on the order of half a single resolution element.
The {\HI} profile is highly suggestive of a narrow plus a broad
component combining to yield a deep core with a broad wing.  The VP
components reflect this profile shape, with cloud \#7 providing the
profile core, $\log N({\HI}) = 14.1$ and $b({\HI}) = 8$~{\kms}, and
with cloud \#8 providing the broadened wings, $\log N({\HI}) = 13.7$
and $b({\HI}) = 43$~{\kms}.  The requirement for the broader cloud is
significant at the 99\% confidence level, even though some
contribution to the blue portion of the broad wing is due to the
extremely broad VP component (cloud \#11) at $v = +76$~{\kms}.

At $v \simeq -120$~{\kms}, {\CIII}, {\CIV} and {\OVI} are formally
detected in the spectra.  However, from the VP fitting, both clouds
\#7 and \#8 have associated {\CIII} and {\OVI}, whereas only cloud \#7
has associated {\CIV}.  Inspection of Figure~\ref{fig:metals} shows
that the {\CIV} absorption appears to be offset in velocity from the
{\OVI}.  However, the {\CIV} is measured in the STIS spectrum, whereas
the {\OVI} and {\CIII} are measured in the COS spectrum.  Thus, the
measured velocity offset relies heavily on the accuracy of the
wavelength zero points between the two independent spectra.

Assuming thermal broadening, the implied temperature of cloud \#7 is
8000~K, which is too cool to host strong {\OVI} and {\CIV} absorption.
The implied temperature of cloud \#8 is $T=114,000\pm 38,000$~K, which
is consistent with the temperature at which {\CIV} absorption peaks
for collisional ionization equilibrium.  To analyze this absorption
structure and constrain its hydrogen density and metallicity, we opted
to assume a single phase of gas that is dominated by cloud \#8
(however, in \S~\ref{sec:hydroeq} we also model these clouds as two
distinct ionization phases).  We added the {\HI} and metal column
densities from clouds \#7 and \#8.  For this exercise, we found that
the column density limits for {\MgII}, {\SiIII}, and {\SiIV} did not
contribute to constraining the gas properties.

In Figure~\ref{fig:cloud78}, we illustrate our analysis of clouds
\#7+8.  The ionization code was run for $-7 \leq \log \hbox{\hden}
\leq -1$ for $T=114,000$~K, and for $T=152,000$ and $T=76,000$, which
brackets the $1~\sigma$ uncertainty in the temperature.  In
Figure~\ref{fig:cloud78}$a$, we illustrate the model column densities
for {\CIII}, {\CIV}, and {\OVI} for $\log N({\HI}) = 14.3$ and
metallicity $\log Z/Z_{\odot} = -2.0$.  In
Figure~\ref{fig:cloud78}$b$, we show the metallicity constraints as a
function of {\hden} for each metal ion, where the range in the
constraints are based on the uncertainty in the temperature and the
uncertainty in the {\HI} and metal ion column densities.  The region
of overlap indicates where the three ions provide consistent
constraints on {\hden} and $\log Z/Z_{\odot}$.  Allowing for the
uncertainties in the data, the gas is constrained to have $-4.2 \leq
\log \hbox{\hden} \leq -3.7$ and $-2.0 \leq \log Z/Z_{\odot} \leq
-1.6$ by all three metal ions.  Note that metallicity is contrained
primarily by $N({\CIII})$, whereas hydrogen density is constrained
primarily by $N({\OVI})$.

\subsection{Cloud \#10}
\label{sec:cld10}

Cloud \#10 is located at $v \simeq -30$~{\kms} with respect to the
systemic velocity of G1.  It is optically thin in neutral hydrogen,
with $\log N({\HI}) = 13.7 \pm 0.2$ and an implied temperature of $T =
20,500 \pm 8600$~K.  The only detected metal line absorption is from
{\OVI}, with $\log N({\OVI}) = 13.6 \pm 0.1$.  

A high abundant the O$^{+5}$ ion in a $T=20,000$~K cloud may be
somewhat surprising.  For $\hbox{\hden} \leq -3$ and $\log Z/Z_{\odot}
= -2$, analysis of the rate coefficients output by the ionization code
show that the balance of O$^{+4}$, O$^{+5}$, and O$^{+6}$ is dominated
by photoionization and recombination via charge exchange with H$^{+}$,
which dominates over free electron recombination by 2 orders of
magnitude for O$^{+5}$.

For cloud \#10, the measured upper limit on {\CIII} is $\log
N({\CIII}) \leq 12.7$ and on {\CIV} is $\log N({\CIV}) \leq 13.2$.  As
with with our exercise for clouds \#7+8, we found that the column
density limits for {\MgII}, {\SiIII}, and {\SiIV} did not contribute
to constraining the gas properties.

In Figure~\ref{fig:cloud10}, we illustrate our analysis of cloud \#10.
The ionization code was run for $-7 \leq \log \hbox{\hden} \leq -1$
for $T=20,000$~K.  In Figure~\ref{fig:cloud10}$a$, we show the column
densities for {\CIII}, {\CIV}, and {\OVI} for $\log N({\HI}) = 13.7
\pm 0.2$ and metallicity $\log Z/Z_{\odot} = -1.4$.  The metallicity
and hydrogen density constraints for cloud \#10 are shown in
Figure~\ref{fig:cloud10}$b$.

If we account for the uncertainty in the implied temperature, we find
that for $T=28,000$~K, the curves in Figure~\ref{fig:cloud10}$b$ are
unchanged for $\log \hbox{\hden} \geq -5.3$, and are shifted upward in
the diagram by no more than 0.7 dex (higher $Z$ at a given {\hden})
for $\log \hbox{\hden} < -5.3$.  As temperature is decreased, the
curves shift to the upper right in the diagram such that at
$T=12,000$~K, a minimum metallicity in the allowed range $-2.3 \leq
\log Z/Z_{\odot} \leq -1.8$ occurs at $\log \hbox{\hden} \geq -3.7$
instead of $-2.5 \leq \log Z/Z_{\odot} \leq -2.0$ at $\log
\hbox{\hden} = -5.3$.  It seems less likely that {\OVI} absorption
would be as strong as we detect in 12,000 degree gas.

\subsection{Clouds with No Detected Metal Lines}

The remainder of the {\HI} absorption complex is characterized by
multiple components with gas temperatures in the range
20,000--90,000~K, though a hot, nearly million degree component is
also present (cloud \#11).  For the clouds with no detected metals,
the densities, ionization corrections, and therefore metallicities
cannot be constrained without additional assumptions.  In
\S~\ref{sec:hydroeq}, we assume hydrodynamic equilibrium and undertake
an analysis of the clouds with only upper limits on the metal line
column densities.  Below, we motivate the assumption of hydrodynamic
equilibrium as a reasonable scenario.

\section{Inferred Physical Conditions of the Gas}
\label{sec:gasconditions}


\subsection{Ionization Equilibrium and Cloud Stability}

In general, ionization equilibrium is valid when the collisional
ionization time scale, $\tau _{\rm coll}$, and the photoionization
time scale, $\tau _{\rm ph}$, are shorter than the cooling time of the
gas, $\tau _{\rm cool}$.

For a monatomic gas, the cooling time is the ratio of the energy
per unit volume and the energy loss per unit volume per unit time,
$\tau_{\rm cool} = Q/\dot{Q} = \sfrac{3}{2}(n_{\rm e}+n_{\hbox{\tiny A}}) \, kT /
n_{\rm e} n_{\hbox{\tiny A}} \Lambda (T,Z)$, where
$n_{\rm e}$ is the electron density, $n_{\hbox{\tiny A}}$ is the total
density of all ions, and $\Lambda (T,Z)$ is the specific cooling
function for metallicity $Z$.  We obtain $n_{\rm e}$ and
$n_{\hbox{\tiny A}}$ from our ionization code as a function of {\hden}
and $T$.  Our cloud models are optically thin, such that
$n_{\hbox{\tiny A}} \propto n_{\hbox{\tiny H}}$ and $n_{\rm e} \propto
n_{\hbox{\tiny H}}$ at a given $T$, we thus find that the behavior of
the cooling time follows $\tau _{\rm cool} \propto n_{\hbox{\tiny
H}}^{-1}$ at fixed $T$.

We adopt the $Z=0$ specific cooling function of \citet{mappings} for a
gas in collisional ionization equilibrium \citep[also
see][]{diffuseuniverse}.  Since our models include photoionization, we
corrected for photoionization heating \citep[see][]{agnsquared}, 
which reduces the magnitude of $\dot{Q}$, and therefore increases the
estimated cooling time.  We found that photoionization heating from
the UVB at $z=0.67$ is negligible for the density and temperatures
ranges we explored.

Metals strongly contribute to the cooling rate, so the $Z=0$ curve
provides an upper limit on the estimated cooling time under the
assumption of ionization equilibrium.  However, the effect is negligible
for $\log Z/Z_{\odot} = -3$ and is a maximum difference of 0.4 dex at
$\log T \simeq 5.3$ for $\log Z/Z_{\odot} = -2$
\citep[see][]{mappings}.

For a given ion, the collisional time scale is well approximated as
$\tau _{\rm coll} \simeq 1/[n_{\rm e}(\alpha_{\rm rec}+ \alpha_{\rm
coll})]$, whereas the photoionization time scale is $\tau _{\rm ph}
\simeq 1/R_{\rm ph}(J_\nu)$, where $\alpha_{\rm rec}(T)$ and
$\alpha_{\rm coll}(T)$ are the recombination and collisional
ionization rate coefficients, and $R_{\rm ph}(J_\nu)$ is the
photoionization rate for the given ion, respectively.  We obtain these
quantities directly from our ionization code.  Since $n_{\rm e}
\propto n_{\hbox{\tiny H}}$ at a given $T$, we have $\tau _{\rm coll}
\propto n_{\hbox{\tiny H}}^{-1}$ at fixed $T$.  Thus, the ratio
$\tau_{\rm cool}/\tau_{\rm coll}$ is effectively constant as a
function of {\hden} for fixed temperature.  Note that for lower
temperatures, $\alpha_{\rm coll}(T)$ vanishes so that $\tau _{\rm
coll}$ becomes equal to the recombination time scale, $\tau _{\rm
rec} \simeq 1/n_{\rm e}\alpha_{\rm rec}$.

\begin{deluxetable*}{lcrrrrrrrrrrrrrc}
\tablecolumns{16}
\tablewidth{0pt}
\tablecaption{Cloud Models of the {\HI} Complex\tablenotemark{a} \label{tab:cldphys}}
\tablehead{
\colhead{Cld\#}              & 
\colhead{$N({\HI})$}    & 
\colhead{$T$}           & 
\colhead{{\hden}}            & 
\colhead{$f_{\hbox{\tiny HI}}$} & 
\colhead{$\alpha_{\rm rec}$}  &
\colhead{$\alpha_{\rm coll}$}  &
\colhead{$\tau_{\rm cool}$}  & 
\colhead{$\tau_{\rm coll}$}     & 
\colhead{$\tau_{\rm dyn}$}   & 
\colhead{$\tau_{\rm sc}$}    & 
\colhead{$L$}                & 
\colhead{$M_g$}                &
\colhead{$L_J$}           &
\colhead{$M_J$}           &
\colhead{Note\tablenotemark{b}}              \\ 
\colhead{\phantom{x}}        & 
\colhead{[cm$^{-2}$]}        & 
\colhead{[K]}                & 
\colhead{[cm$^{-3}$]}        & 
\colhead{\phantom{x}}        & 
\colhead{[cm$^{3}$ s$^{-1}$]}  & 
\colhead{[cm$^{3}$ s$^{-1}$]}  & 
\colhead{[yr]}               & 
\colhead{[yr]}               & 
\colhead{[yr]}               & 
\colhead{[yr]}               & 
\colhead{[kpc]}              & 
\colhead{[M$_\odot$]}        &
\colhead{[kpc]}              & 
\colhead{[M$_\odot$]}        &
\colhead{\phantom{x}}        } 
\startdata
  7+8 & 14.30 &  5.06 & $-5.0$ & $-6.62$ &$-13.21$ & $-8.28$ &  9.40 &  5.70 & 10.01 & 11.71 &  4.43 & 15.84 &  2.73 & 10.72 & JU \\
  7+8 & 14.30 &  5.06 & $-4.0$ & $-5.69$ &$-13.21$ & $-8.28$ &  8.40 &  4.70 &  9.51 &  9.78 &  2.51 & 11.05 &  2.23 & 10.22 & JU \\
  7+8 & 14.30 &  5.06 & $-3.0$ & $-5.10$ &$-13.21$ & $-8.28$ &  7.40 &  3.70 &  9.01 &  8.19 &  0.91 &  7.27 &  1.73 &  9.72 & EE \\[5pt]
   10 & 13.70 &  4.31 & $-5.0$ & $-6.02$ &$-12.62$ &-$11.36$ &  8.13 &  8.75 & 10.01 & 10.88 &  3.23 & 12.22 &  2.36 &  9.60 & JU \\
   10 & 13.70 &  4.31 & $-4.0$ & $-5.02$ &$-12.62$ &-$11.36$ &  7.13 &  7.76 &  9.51 &  8.88 &  1.23 &  7.22 &  1.86 &  9.10 & EE \\
   10 & 13.70 &  4.31 & $-3.0$ & $-4.03$ &$-12.62$ &-$11.36$ &  6.14 &  6.77 &  9.01 &  6.89 &$-0.76$&  2.25 &  1.36 &  8.60 & EE \\[5pt]
   11 & 13.43 &  5.68 & $-5.0$ & $-7.22$ &$-13.78$ & $-7.57$ & 10.86 &  4.99 & 10.01 & 11.13 &  4.17 & 15.03 &  3.04 & 11.65 & JU \\
   11 & 13.43 &  5.68 & $-4.0$ & $-6.48$ &$-13.78$ & $-7.57$ &  9.86 &  3.99 &  9.51 &  9.39 &  2.43 & 10.81 &  2.54 & 11.15 & EE \\
   11 & 13.43 &  5.68 & $-3.0$ & $-6.23$ &$-13.78$ & $-7.57$ &  8.86 &  2.99 &  9.01 &  8.14 &  1.17 &  8.04 &  2.04 & 10.65 & EE \\[5pt]
   15 & 13.68 &  4.95 & $-5.0$ & $-6.53$ &$-13.12$ & $-8.50$ &  9.06 &  5.92 & 10.01 & 11.05 &  3.72 & 13.69 &  2.68 & 10.56 & JU \\
   15 & 13.68 &  4.95 & $-4.0$ & $-5.58$ &$-13.12$ & $-8.50$ &  8.06 &  4.92 &  9.51 &  9.10 &  1.77 &  8.84 &  2.18 & 10.06 & EE \\
   15 & 13.68 &  4.95 & $-3.0$ & $-4.88$ &$-13.12$ & $-8.50$ &  7.06 &  3.93 &  9.01 &  7.40 &  0.07 &  4.73 &  1.68 &  9.56 & EE \\[5pt]
   18 & 14.26 &  4.84 & $-5.0$ & $-6.43$ &$-13.03$ & $-8.78$ &  9.06 &  6.20 & 10.01 & 11.59 &  4.20 & 15.14 &  2.62 & 10.40 & JU \\
   18 & 14.26 &  4.84 & $-4.0$ & $-5.46$ &$-13.03$ & $-8.78$ &  8.06 &  5.20 &  9.51 &  9.62 &  2.23 & 10.23 &  2.12 &  9.90 & JU \\
   18 & 14.26 &  4.84 & $-3.0$ & $-4.66$ &$-13.03$ & $-8.78$ &  7.06 &  4.21 &  9.01 &  7.82 &  0.43 &  5.82 &  1.62 &  9.40 & EE \\[5pt]
   20 & 13.10 &  4.41 & $-5.0$ & $-6.09$ &$-12.69$ &-$10.64$ &  8.41 &  8.05 & 10.01 & 10.30 &  2.70 & 10.63 &  2.41 &  9.75 & JU \\
   20 & 13.10 &  4.41 & $-4.0$ & $-5.09$ &$-12.69$ &-$10.64$ &  7.41 &  7.05 &  9.51 &  8.30 &  0.70 &  5.63 &  1.91 &  9.25 & EE \\
   20 & 13.10 &  4.41 & $-3.0$ & $-4.11$ &$-12.69$ &-$10.64$ &  6.41 &  6.07 &  9.01 &  6.32 &$-1.28$&  0.69 &  1.41 &  8.75 & EE 
\enddata
\tablenotetext{a}{All quantities are base 10 logarithmic. The columns
are: (1) cloud number; (2) hydrogen column density, $\log N({\HI})$;
(3) temperature, $\log T$; (4) hydrogen density, $\log \hbox{\hden}$;
(5) hydrogen ionization fraction, $\log f_{\hbox{\tiny HI}}$; (6)
recombination rate coefficient, $\log \alpha _{\rm rec}$; (7)
collisional ionization rate coefficient, $\log \alpha _{\rm rec}$; (8)
cloud cooling timescale, $\log \tau_{\rm cool}$; (9) hydrogen
collisional ionization time scale, $\log \tau_{\rm coll}$; (10) cloud
dynamical time scale, $\log \tau_{\rm dyn}$, (11) cloud sound crossing
time, $\log \tau_{\rm sc}$; (12) cloud absorption length scale, $\log
L$; (13) cloud gas mass, $\log M_g$; (14) cloud Jean's length, $\log L_J$,
and (15) cloud Jean's gas mass, $\log M_J$.}
\tablenotetext{b}{JU denotes that the cloud is Jean's unstable or
moderately Jean's unstable; EE denotes instability or slight
instability toward expansion and/or evaporation.}
\end{deluxetable*}

\subsection{Equilibrium Based On Hydrogen}

In what follows, we briefly explore selected clouds using
representative {\hden} densities for the purpose of illustration.  In
Table~\ref{tab:cldphys}, we present a summary of cloud models for the
{\hden} values $\log \hbox{\hden} = -5$, $-4$, and $-3$, where ion
specific quantities apply for hydrogen. The hydrogen photoionization rate 
from the UVB at $z=0.67$ is $R_{\rm ph} = 3.0 \times
10^{-12}$ s$^{-1}$, yielding $\tau _{\rm ph} = 1 \times 10^4$ yr.  As
we shall see, the photoionization time is significantly shorter than
the cooling times for all models; as such, only the longer collisional
time scales constrain whether the clouds are in ionization
equilibrium.

For clouds \#7+8, at $T=114,000$~K, $\alpha_{\rm rec}= 6.2 \times
10^{-14}$ and $\alpha_{\rm coll} = 5.3 \times 10^{-9}$
cm$^{3}$~s$^{-1}$, respectively.  Assuming $\log
\hbox{\hden} = -4$, we obtain $\tau _{\rm cool} \simeq 3 \times
10^{8}$ yr and $\tau _{\rm coll} = 5 \times 10^4$ yr.  With $\tau
_{\rm cool}/\tau _{\rm coll} \simeq 5\times 10^{3}$, hydrogen is
clearly in ionization equilibrium in clouds \#7+8; the short
photoionization and collisional ionization time scales relative to the
recombination time scale are indicative of the highly ionized
condition ($f_{\hbox{\tiny HI}} = 2 \times 10^{-6}$).  The condition
$\tau _{\rm ph} \la \tau _{\rm coll}$ indicates photoionization
marginally dominates over collisional ionization.

For cloud \#10, at $T=20,000$~K, $\alpha_{\rm rec} = 2.4 \times
10^{-13}$ and $\alpha_{\rm coll} = 4.3 \times 10^{-12}$
cm$^{3}$~s$^{-1}$.  This cloud is constrained to have $\log
\hbox{\hden} < -4$; assuming $\log \hbox{\hden} = -5$, we obtain $\tau
_{\rm cool} \simeq 1 \times 10^{8}$ yr and $\tau _{\rm coll} = 6
\times 10^8$ yr.  For this assumed density, we find $\tau _{\rm
cool}/\tau _{\rm coll} \simeq 0.2$, suggesting that ionization
equilibrium may be marginal in cloud \#10 for hydrogen (the gas
temperature evolves on a similar time scale that the balance can be
achieved).  As listed in Table~\ref{tab:cldphys}, the thermal time
scale is shorter than the ionization time scale for the presented
range of {\hden}.

For cloud \#11, the hot, $\log T = 5.68$~K, cloud, we have
$\alpha_{\rm rec} = 1.7 \times 10^{-14}$ and $\alpha_{\rm coll} = 2.7
\times 10^{-8}$ cm$^{3}$~s$^{-1}$.  For $\log \hbox{\hden} =-4$ and
$-5$, we obtain $\tau _{\rm cool} \simeq 7 \times 10^{9-10}$ yr, and
$\tau _{\rm coll} = 1 \times 10^{4-5}$ yr, where the longer times
correspond to the lower density.  Cloud \#11 has $\tau _{\rm
cool}/\tau _{\rm coll} \simeq 7 \times 10^{5}$, indicating an
ionization equilibrium condition.  Note that if $\log \hbox{\hden}
=-4$, we have $\tau_{\rm ph} \simeq \tau_{\rm coll}$, indicating that
hydrogen ionization ($f_{\hbox{\tiny HI}} = 3 \times 10^{-7}$) is
driven by both photo and collisional processes.  If $\log \hbox{\hden}
=-5$, then $f_{\hbox{\tiny HI}} = 6 \times 10^{-8}$, which is
dominated by photoionization ($\tau _{\rm coll}/\tau _{\rm ph} = 10$);
the larger ratio due to the longer recombination time scale results in
a higher ionization condition for this lower density.

With regard to the remaining clouds in the {\HI} complex with only
limits on the metal line measurements, we selected clouds \#15 [$\log
N{(\HI}) = 13.7, \log T = 4.95$], \#18 [$\log N{(\HI}) = 14.3, \log T
= 4.84$], and \#20 [$\log N{(\HI}) = 13.1, \log T = 4.41$] as
representative.  Of clouds with no detected metals, \#15 has the
highest temperature and an intermediate $N({\HI})$ as compared to all
clouds in the {\HI} complex.  Cloud \#20 has the lowest temperature
and the smallest $N({\HI})$, and cloud \#18 has an intermediate
temperature and the highest $N({\HI})$.  The models, for hydrogen, are
given in Table~\ref{tab:cldphys}.  Note that $\tau_{\rm cool} >
\tau_{\rm coll}$ for these representative clouds, indicating the
validity of ionization equilibrium.

\subsection{Equilibrium Based On O$^{+5}$ and C$^{+3}$}

For strict ionization equilibrium to hold, the ionization time scales
must be smaller than the cooling time for all ions.  The same
arguments invoked for hydrogen above apply to the metals.  Consider
the O$^{+5}$ ion; the photoionization rate from the UVB at $z=0.67$ is
$R_{\rm ph} = 4.3 \times 10^{-16}$ s$^{-1}$, which yields $\tau_{\rm
ph} = 7 \times 10^7$ yr.

For clouds \#7+8, at $T=114,000$~K, $\alpha_{\rm rec}$ and
$\alpha_{\rm coll}$ are $3.7 \times 10^{-12}$ and $3.8 \times
10^{-16}$ cm$^{3}$~s$^{-1}$, respectively.  For $\log \hbox{\hden} =
-4$, we obtain $\tau _{\rm coll} = \tau _{\rm rec} = 7 \times 10^7$
yr, where both scale in proportion to $n_{\hbox{\tiny H}}^{-1}$.  We
thus have $\tau_{\rm coll} = \tau_{\rm rec} \simeq 4 \, \tau_{\rm
cool}$ independent of {\hden}, indicating that the O$^{+5}$ ion is
marginally in ionization equilibrium.  Similarly, we find $\tau_{\rm
coll} \simeq 3 \, \tau_{\rm cool}$ for C$^{+3}$.

For cloud \#10, at $T=20,000$~K, $\alpha_{\rm rec} = 1.4 \times
10^{-11}$ cm$^{3}$~s$^{-1}$ and $\alpha_{\rm coll}$ is negligible,
indicating that collisional processes have vanishing importance in the
ionization balance of O$^{+5}$.  For $\log \hbox{\hden} = -5$, we
obtain $\tau_{\rm coll} = \tau _{\rm rec} = 2 \times 10^8$ yr.  For
all densities, we have $\tau _{\rm cool}/\tau _{\rm coll} \simeq 0.8$.
Similarly, for C$^{+3}$, we find $\tau _{\rm cool}/\tau _{\rm coll}
\simeq 0.3$.  As with hydrogen, ionization equilibrium may be marginal
for cloud \#10 for O$^{+5}$ and C$^{+3}$.

For cloud \#11, at $\log T=5.68$, we have $\alpha_{\rm rec} = 1.2
\times 10^{-12}$ and $\alpha_{\rm coll} = 6.9 \times 10^{-11}$
cm$^{3}$~s$^{-1}$ for O$^{+5}$.  When $\log \hbox{\hden} =-4$, and
$-5$, we obtain $\tau _{\rm coll} = 4 \times 10^{6-7}$ yr, where the
longer times correspond to the lower density.  We have $\tau _{\rm
cool}/\tau _{\rm coll} \simeq 2000$ for which O$^{+5}$ is
predominantly in collisional equilibrium, i.e., $\tau_{\rm
coll}/\tau_{\rm ph} \simeq 20$ if $\log \hbox{\hden} = -4$, and
roughly equally balanced by photo and collisional equilibrium, i.e.,
$\tau_{\rm coll} \simeq \tau_{\rm ph}$ for $\log \hbox{\hden} = -5$.
For C$^{+3}$, we obtain $\tau _{\rm cool}/\tau _{\rm coll} \simeq 2
\times 10^{4}$.  Cloud \#11 is clearly in ionization equilibrium.

For clouds \#15, \#18, and \#20, we find $\tau _{\rm cool}/\tau _{\rm
coll} \simeq 2$, $2$, and $1$, respectively for O$^{+5}$.  For
C$^{+3}$, we find $\tau _{\rm cool}/\tau _{\rm coll} \simeq 0.9$,
$0.9$, and $0.4$, respectively.  For these clouds, O$^{+5}$ is
marginally in ionization equilibrium, whereas C$^{+3}$ is marginally
not in equilibrium.

Note that the cooling times are 1--3 orders of
magnitude shorter than the Hubble time, $\tau_{\hbox{\tiny H}} = 1/H_0
= 1.4 \times 10^{10}$ yr, except for cloud \#11 for $\log
\hbox{\hden} \leq -5$ (recall that the cooling time scales inversely
with {\hden} for fixed $T$).  This would indicate that the cloud
thermal conditions have evolved; however it is not possible to
estimate the thermal histories of the clouds beyond speculating that
they were hotter (and presumably less dense) at $z > 0.67$.  For $ 4
\leq \log T \leq 4.7$ and for $5 \leq \log T \leq 6$, the cooling time
increases toward higher temperatures, so we can infer that the thermal
evolution of clouds in these temperature was slower at epochs prior to
$z=0.67$ assuming the cloud densities have not strongly evolved.

\subsection{Dynamical Stability, Sizes, and Masses}

In view of the inferred thermal evolution of the absorbing clouds, it
is interesting to examine their dynamical stability.  This can be
achieved by comparing the cloud dynamical and sound crossing time
scales.  Additionally, it would be of interest to estimate the cloud
sizes and masses.  

To within a dimensionless factor of order unity, the dynamical time
\citep[see][\S~14.2.1]{lequeux}, or free-fall time, of a cloud with
total mass density $\rho = \rho_g/f_g = (A_{\hbox{\tiny H}} m_{\rm a}
/ x_{\hbox{\tiny H}} f_g ) \, \hbox{\hden} $, is $\tau_{\rm dyn} =
1/\sqrt{G\rho}$, where $A_{\hbox{\tiny H}}$ is the atomic weight of
hydrogen, $m_{\rm a}$ is the atomic mass unit, $x_{\hbox{\tiny H}} =
0.72$ is the mass fraction of hydrogen\footnote{We have assumed a
solar helium to hydrogen abundance ratio and solar metal abundance
ratios for $\log Z/Z_{\odot} = -2$.  A low metallicity primordial
abundance pattern has $x_{\hbox{\tiny H}} \simeq 0.76$, which yields a
5\% difference in $\rho_g$.}  and $f_g = \Omega_b/\Omega_m \simeq
0.16$ is the fraction of the baryonic mass in gas.  Evaluating, we
have $\tau_{\rm dyn} \simeq 3.2 \times 10^7 \hbox{\hden}^{-1/2}$ yr.

The sound crossing time is $\tau_{\rm sc} = L/c_s$, where $L$ is the
physical length scale of the cloud and the sound speed is $c_s^2 =
\gamma P_g/\rho_g = \gamma kT/\mu m_{\rm a}$, where $P_g$ is the gas
pressure, $\gamma =5/3$ for an ideal monatomic gas, and $\mu$ is the
mean molecular weight.  Our ionization models indicate that the clouds
are highly ionized, so we assume $\mu = (2x_{\hbox{\tiny H}} +
\sfrac{3}{4} x_{\hbox{\tiny He}} + \sfrac{1}{2}x_{\hbox{\tiny Z}}
)^{-1} \simeq 0.6$, appropriate for a fully ionized gas with $\log
Z/Z_{\odot} \simeq -2$.  We thus obtain, $\tau_{\rm sc} \simeq
6.4\times 10^{9} \, L \, T^{-1/2}$ yr, when $L$ is given in kpc.

The ``absorption'' length scale of the cloud can be estimated as the
line of sight path length required to give rise to the measured
$N({\HI})$ for the inferred {\hden} from the ionization models, $L =
N({\HI})/(f_{\hbox{\tiny HI}}\hbox{\hden})$.  Assuming spherical
geometry, the gas mass of the cloud is crudely estimated from $M_g =
\rho_g L^3 \simeq 3.4 \times \! 10^7 \hbox{\hden} L^3$~M$_{\odot}$
when $L$ is given in kpc.  A spherical geometry with unity volume
filling factor is likely a very poor model, and one which will
significantly overestimate the cloud mass.  If the absorber geometry
is cylindrical of length $L$ and radius $R$ with aspect ratio $\beta =
2R/L$, then $M_{\rm cyl}/M_{\rm sph} \propto \beta^2$ if the line of
sight probes parallel to $L$.

It is well established based on observational and theoretical grounds,
that in general, {\Lya} forest clouds cannot be pressure confined
\citep{rauch98}.  \citet{schaye01} convincingly argues that clouds
which develop due to the gravitational influence of underlying dark
matter density perturbations persist in a near hydrodynamic
equilibrium state {\it local to the region giving rise to the {\Lya}
absorption}.  Effectively, this is equivalent to stating that the
dynamical time is equal to the sound crossing time, $\tau _{\rm dyn}
\simeq \tau_{\rm sc}$, consequently implying that the characteristic
length of the {\it absorbing region\/} will be on the order of the
local Jean's length, $L_J$ \citep[see][\S~14.1.2]{lequeux}.  Equating
$\tau _{\rm dyn}$ and $\tau_{\rm sc}$, the Jean's length is $L_J
\simeq 5.0 \times 10^{-3} \, (T/\hbox{\hden})^{1/2}$ kpc, from which
the Jean's gas mass\footnote{The Jean's mass, $M_J$, usually applies
to the total mass.  If the gas fraction, $f_g = \Omega_b/\Omega_m$, in
these clouds is near the cosmic mean, then gas mass and total mass are
related by $M_{\rm tot} = M_g/f_g$.}  can be estimated, $M_J = \rho_g
L_J^3 \simeq 4.3 \, T^{3/2} \hbox{\hden}^{-1/2}$~M$_{\odot}$.

For $M_g > M_J$ the condition is $\tau_{\rm dyn} < \tau_{\rm sc}$; the
cloud is Jean's unstable to further gravitational contraction and will
adjust on the dynamical time scale via fragmentation or due to shock
processes.  Conversely, for $M_g < M_J$ the condition is $\tau_{\rm
sc} < \tau_{\rm dyn}$, and the cloud will adjust in the sound
crossing time scale, either via evaporation or expansion
\citep{schaye01,lequeux}.  The criterion assumes that the cloud is
isolated, spherical, homogeneous, and exhibits no bulk motions.

The dynamical and sound crossing times, deduced cloud sizes and cloud
gas masses are listed in Table~\ref{tab:cldphys} for cloud \#7+8,
\#10, \#11, and the representative clouds \#15, \#18, and \#20 for
$\log \hbox{\hden} = -3$, $-4$, and $-5$.  As described above, cloud
\#7+8 is well constrained to have $\log \hbox{\hden} \simeq -4$,
whereas cloud \#10 and the remaining clouds with only limits on the
metal line column densities are not well constrained.  

For $ \log \hbox{\hden} \leq -3$, $-4$, and $-5$, we obtain $\tau_{\rm
dyn} \simeq 1 \times 10^9$, $3 \times 10^{9}$, and $1 \times
10^{10}$~yr, respectively.  Note that $\tau _{\rm dyn} \propto
\hbox{\hden}^{-1}$, whereas $\tau _{\rm sc} \propto (f_{\hbox{\tiny
HI}}\hbox{\hden})^{-1}$, and $\tau _{\rm cool} \propto
\hbox{\hden}^{-1/2}$.  The relative behavior of these times scales is
such that a transition from $\tau_{\rm sc} < \tau_{\rm dyn}$ to
$\tau_{\rm sc} > \tau_{\rm dyn}$ occurs as {\hden} increases.  This
indicates that for larger {\hden}, the clouds have a propensity to be
in the regime of Jean's instability, whereas as for smaller {\hden},
the clouds become unstable to evaporation and expansion.  The last
column of Table~\ref{tab:cldphys} notes whether a cloud model is
inferred to be Jean's unstable (JU) or unstable to expansion and/or
evaporation (EE).  The behavior with {\hden} suggests that between $-3
\leq \log \hbox{\hden} \leq -5$ there is a density at which the clouds
would classify as being in hydrodynamic equilibrium.  This is also
reflected in the behavior of the Jean's lengths, $L_J$, which would be
equivalent to the absorption length scales, $L$, when $\tau_{\rm dyn}
= \tau_{\rm sc}$.

Note that the absorption scale lengths, $L$, and the cloud gas masses,
$M_g$, become unphysically large as hydrogen density decreases.  This
would suggest that the clouds have $\log \hbox{\hden} > -5$, though we
caution that the masses are probably overestimates due to the
assumption of a spherical geometry.  If more akin to cylindrical
structures, viewed along the long axis, a factor of $\beta = 0.1$
reduces the masses by two orders of magnitude.  However, the
absorption length scale along the line of sight is independent of
geometry.

\subsection{Hydrodynamic Equilibrium Conditions}
\label{sec:hydroeq}

\begin{figure*}[thb]
\epsscale{0.8}
\plotone{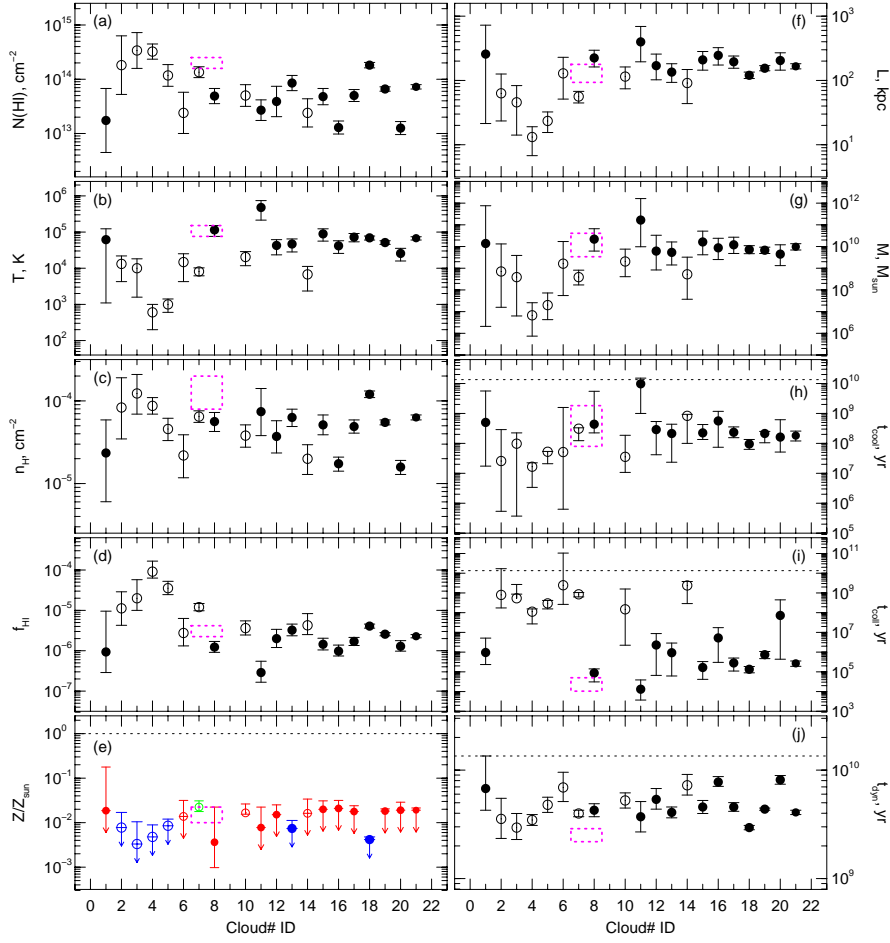}
\caption{The deduced cloud physical conditions under the assumption of
hydrodynamic equilibrium.  --- ($a$) The measured $N({\HI})$ from the
VP fits.  --- ($b$) The measured temperature, based upon the VP
Doppler $b$ parameter. --- ($c$) The equilibrium {\hden} computed from
Eq.~\ref{eq:hydroeq}.  --- ($d$) The equilibrium hydrogen ionization
fraction, $f_{\hbox{\tiny HI}}$, computed from
Eq.~\ref{eq:hydroeq}. --- ($e$) The metallicity, $Z/Z_{\odot}$,
computed from Eq.~\ref{eq:metaleq}; the points are colored coded based
upon the ion that provides the most stringent constraints; red for
O$^{+5}$, green for C$^{+3}$, and blue for C$^{+2}$.  The dotted line
is solar.  --- ($f$) The absorption scale length, $L =
N({\HI})/(f_{\hbox{\tiny HI}}\hbox{\hden})$.  Note that this is
equivalent to the Jean's length, $L_J$, for the assumption of
hydrodynamic equilibrium.  --- ($g$) The gas mass, $M_g$.  Note that
$M_g = M_J$, the Jean's gas mass, for the assumption of hydrodynamic
equilibrium.  --- ($h$) The cooling time, $\tau_{\rm cool}$; the
dotted line is the Hubble time, $\tau_{\hbox{\tiny H}} = 1/H_0$.  ---
($i$) The collisional time, $\tau_{\rm coll}$; the dotted line is
$\tau_{\hbox{\tiny H}}$.  --- ($j$) The dynamical time, $t_{\rm dyn}$;
the dotted line is $\tau_{\hbox{\tiny H}}$.  Note that the sound
crossing time, $\tau_{\rm sc}$ is equivalent to $\tau_{\rm dyn}$ for
the assumption of hydrodynamic equilibrium.  Open data points
represent clouds that do not satisfy the criterion of hydrogen
ionization equilibrium, where solid data points do satisfy the
criterion.  Dotted magenta boxes straddling clouds \#7 and \#8
represent the single-phase model presented in \S~\ref{sec:clds78} as
illustrated in Figure~\ref{fig:cloud78}.}
\label{fig:equilib}
\end{figure*}

In view of the arguments given by \citet{schaye01} that {\Lya} clouds
should be in the regime of hydrostatic equilibrium, and in view of the
unphysically large cloud scale lengths we deduce in the regime of
Jean's instability, it is reasonable to explore the inferred cloud
conditions under the assumption of hydrodynamic equilibrium.  This
assumption will yield an equilibrium value for {\hden}, from which the
ionization conditions and metallicities can be constrained.

Equating the dynamical time and the sound crossing time, and invoking
the absorption scale length, $L = N({\HI})/(f_{\hbox{\tiny
HI}}\hbox{\hden})$, we derive the condition of hydrodynamic
equilibrium,
\begin{equation}
n_{\hbox{\tiny H}} \, x_{\hbox{\tiny H}} f_g \, \gamma kT \,
[f_{\hbox{\tiny HI}}(n_{\hbox{\tiny H}},T)]^2
= 
G A_{\hbox{\tiny H}} \mu m^2_{\rm a} 
  \, [N({\HI})]^2  \, ,
\label{eq:hydroeq}
\end{equation}
where $N({\HI})$ and $T$ are measured from the VP fit models to the
{\Lya} profiles, and $f_{\hbox{\tiny HI}}(\hbox{\hden},T)$ and {\hden}
are computed using ionization models of the clouds.

Since the hydrogen ionization fraction is a function of
$n_{\hbox{\tiny H}}$, Eq.~\ref{eq:hydroeq} is a transcendental
equation and must therefore be solved numerically for the equilibrium
hydrogen density for a cloud with temperature $T$.  For each cloud, we
interpolate a grid of ionization models with metallicity $\log Z_{\rm
m}/Z_{\odot}=-2$ for $-6.0 \leq \log \hbox{\hden} \leq -2.0$ in steps
of $\Delta \log \hbox{\hden} = 0.2$ and $2.6 \leq \log T \leq 6.0$ in
steps of $\Delta \log T = 0.1$.  We set $T$ equal to the measured
cloud temperature (based upon the Doppler $b$ parameter from the VP
fits), interpolate to obtain $f_{\hbox{\tiny HI}}(\hbox{\hden},T)$ as
a function {\hden} at that $T$, and iterate using Brent's method to
locate the $n_{\hbox{\tiny H}}$ that satisfies Eq.~\ref{eq:hydroeq} to
a tolerance $1\times 10^{-10}$.  The method assumes that
$f_{\hbox{\tiny HI}}(\hbox{\hden},T)$ is independent of metallicity
(which we verified for $\log Z_m/Z_{\odot} \leq -1$).

For this exercise, we treat clouds \#7 and \#8 separately (i.e., as a
multiphase structure as opposed to a single-phase cloud as done in
\S~\ref{sec:clds78}).  We also omit cloud \#9, which has a very
uncertain Doppler $b$ parameter in that its temperature is consistent
with $T=0$ K.

Once Eq.~\ref{eq:hydroeq} is satisfied, the cloud absorption scale
length, $L$, can be computed, from which the cloud gas mass can be
estimated.  Note that these quantities, while determined from the
measured $N({\HI})$ and $T$ as constrained by the ionization models,
will be equivalent to the Jean's length and Jean's gas mass for the
equilibrium $n_{\hbox{\tiny H}}$.

For each cloud, the uncertainty in the equilibrium $n_{\hbox{\tiny
H}}$ and $f_{\hbox{\tiny HI}}$ are obtained by accounting for the
measured uncertainties in both $N({\HI})$ and $T$ from the VP fits.
The uncertainties in $L$ account for the uncertainties in $N({\HI})$,
{\hden}, and $f_{\hbox{\tiny HI}}$. The uncertainties in $M_g$ account
for the uncertainties in $\rho_g$ (due to $n_{\hbox{\tiny H}}$) and
$L$.

Once the equilibrium values are determined, the cloud metallicities
and their uncertainties can be estimated from the equilibrium
ionization cloud model.  Denoting this metallicity as $Z$ (in
solar units), we have
\begin{equation}
\log Z = \log \left[ \frac{N({\rm
X})}{N({\HI})} \right] _{\hbox{\tiny VP}} + \log \left[
\frac{Z_{\rm m} f_{\hbox{\tiny HI}}\hbox{
\hden}}{n_{\hbox{\tiny X}}} \right]
_{\rm eq} \, ,
\label{eq:metaleq}
\end{equation}
where $N({\HI})$ is the measured neutral hydrogen column density
obtained from the VP fit to the data, $N({\rm X})$ is the measured or
upper limit on the column density species X from the VP modeling, and
$n_{\hbox{\tiny X}}$ is the number density of species X from the
ionization model for the equilibrium $f_{\hbox{\tiny HI}}$ and {\hden}
and the model metallicity $Z_{\rm m}$.  That is, we scale the
equilibrium ionization model with $Z_{\rm m}$ to obtain $Z$
and an estimate of its uncertainty.

In Figure~\ref{fig:equilib}, we present the deduced cloud properties
under the assumption of hydrodynamic and ionization equilibrium.
Solid data points satisfy the condition of hydrogen ionization
equilibrium, whereas open data points do not.  The measured $N({\HI})$
and $T$ are shown in panels \ref{fig:equilib}$a$ and
\ref{fig:equilib}$b$, respectively.  The equilibrium {\hden} and
$f_{\hbox{\tiny HI}}$ are presented in panels \ref{fig:equilib}$c$ and
\ref{fig:equilib}$d$, respectively.  Note that, as suggested by the
analysis presented in Table~\ref{tab:cldphys}, the clouds have $-4.8
\leq \log \hbox{\hden} \leq -3.8$ and $ -6 \leq \log f_{\hbox{\tiny
HI}} \leq -4$.

The metallicities are plotted in Figure~\ref{fig:equilib}$e$. Except
for clouds \#7, \#8, and \#10, the metallicities are upper limits.
For the assumption of single ionization phase clouds, each metal line
provides a unique limit.  The most stringent limits are presented,
with the data point color coded by the ion that provides this best
limit (red for O$^{+5}$, green for C$^{+3}$, and blue for C$^{+2}$).
In general, the upper limits are $\log Z/Z_{\odot} < -1.7$.  The
measured values for clouds \#7, \#8, and \#10 are $\log Z/Z_{\odot} =
-1.65^{+0.14}_{-0.10}$, $-2.44^{+0.79}_{-0.57}$, and
$-1.78^{+0.20}_{-0.03}$, respectively.

The absorption scale length (which is equal to the Jean's length under
the assumption of hydrodynamic equilibrium), is shown in
Figure~\ref{fig:equilib}$h$ for each cloud.  The typical scale length
is few hundred kpc, except for clouds \#2, \#3, \#4, and \#5, which
have the lowest temperatures and thus relatively high densities and
low ionization conditions.  Recall, that the VP fits for clouds \#1
through \#6 should be viewed with caution.  The gas masses, $M_g$, of
the clouds (which equal the Jean's gas masses under the assumption of
hydrodynamic equilibrium) are presented in
Figure~\ref{fig:equilib}$g$.  Assuming spherical clouds, the gas
masses are on the order of $10^{8}$-$10^{10}$~M$_{\odot}$, except for
cloud \#11, which has $M_g \simeq 10^{11}$~M$_{\odot}$.  These mass
estimates should be viewed as upper limits by as much as one to two
orders of magnitude.  Cloud \#11 is the highest temperature cloud with
the highest ionization condition.  The larger mass is a result of the
large scale length and the fact that the hydrogen density is
relatively high, $\log \hbox{\hden} \simeq -4.1$.

The cooling time, $\tau_{\rm cool}$, is presented in panel
\ref{fig:equilib}$h$.  Note that under the assumption of hydrodynamic
equilibrium, the clouds have thermal stability in the order of 1--2
orders of magnitude shorter than the Hubble time.  The collisional
time, $\tau_{\rm coll}$, is presented in panel \ref{fig:equilib}$i$.
Except for clouds \#10 and \#14, the clouds are in ionization
equilibrium.  However, clouds \#2 through \#6 also may not be in
ionization equilibrium; we again remind the reader that the VP fits to
these latter clouds are to be viewed with caution.  The dynamical time
(which is equal to the sound crossing time under the assumption of
hydrodynamic equilibrium), is presented in Figure~\ref{fig:equilib}$j$
for each cloud.  In all cases, $\tau_{\rm dyn}$ and $\tau_{\rm sc}$
are a factor of a few less than the Hubble time, which is indicated by
the dashed line.

The values obtained from this exercise are in remarkable agreement
with those predicted from the simple scaling relations proposed by
\citet{schaye01} for the assumption of hydrodynamic equilibrium.

For comparison between the individual models of clouds \#7 and \#8 and
the combined cloud \#7+8 modeled in \S~\ref{sec:clds78}, we indicate
the results of the latter analysis as dashed boxes on
Figure~\ref{fig:equilib}.  For cloud \#7+8, the measured column
densities were summed, from which the density and metallicity were
simultaneously constrained assuming a single ionization phase (see
Figure~\ref{fig:cloud78}).  That analysis resulted in a slightly
larger {\hden} than the assumption of hydrodynamic equilibrium in the
individual clouds, through the metallicities are consistent between
analysis methods.  The difference in {\hden} is likely due to the
adding of the column densities.  We also compare the individual cloud
equilibrium and the combined cloud $L$, $M_g$, $\tau_{\rm cool}$, and
$\tau_{\rm dyn}$ $\tau_{\rm coll}$, shown as the dashed boxes on
panels \ref{fig:equilib}$f$--$j$.  Note that cloud \#7 has $\tau_{\rm
coll} \gg \tau_{\rm ph}$, where $\tau_{\rm ph} = 10^4$ yr,
indicating that the hydrogen in this cooler cloud is photoionized.  On
the other hand, cloud \#8 has $\tau_{\rm coll} \simeq \tau_{\rm ph}$
for both hydrogen and O$^{+5}$, and thus has a substantial collisional
ionization contribution.  

Cloud \#11 is of particular interest.  This component is the hottest
and most highly ionized, $f_{\hbox{\tiny HI}} \simeq 3 \times
10^{-7}$, cloud in the complex, and is among the highest density,
$\log n_{\hbox{\tiny H}} \simeq -4.1$, of the clouds.  We find
$\tau_{\rm coll}/\tau_{\rm ph} \simeq 1 $ for hydrogen, and $\tau_{\rm
coll}/\tau_{\rm ph} < 100 $ for O$^{+5}$; thus hydrogen is equally
photo and collisionally ionized, whereas O$^{+5}$, though not detected,
is predominantly collisionally ionized.  It is plausible that this
cloud (\#11) is shock heated gas, as further suggested by the fact
that the deduced cooling time ($\tau _{\rm cool} \simeq 10$ Gyr) is
substantially longer than the deduced dynamical time ($\tau_ {\rm dyn}
\simeq 4 $ Gyr).  Crudely adopting the dynamical time as a proxy for
the compression time \citep[e.g.,][]{birnboim03,dekel06}, and
considering the temperature and ionization conditions, we find that
cloud \#11 is the only component in the {\HI} complex that is
suggestive of shocked gas.

\subsection{Caveats}

The analysis we have presented has employed many simplifying
assumptions.  The VP fitting method philosophy assumes that the gas
structure comprises several spatially distinct isothermal clouds.  In
fact, it is very possible that the {\HI} complex is a quasi-continuous
nonuniform structure with temperature and density variations having a
range of bulk motions (and possibly at least one shock front, i.e.,
cloud \#11).  It is also plausible that such bulk motions can align in
line of sight velocities creating caustics that emulate distinct
clouds so that some VP components actually model a heterogeneous
physical condition.

Furthermore, the analysis invoking the dynamical time, sound crossing
time, and Jean's mass and length is predicated on a homogeneous cloud.
A cloud in thermodynamic equilibrium cannot simultaneously be
homogeneous and isothermal, as we have assumed here.  

In support of the assumption of hydrodynamic equilibrium, we note that
\citet{schaye01} argues that a spherical cloud with an isothermal
density profile, i.e., $n_{\hbox{\tiny H}}(r) \propto r^{-2}$, has a
well defined characteristic {\hden} that is on the order of the maximum
density probed by the line of sight.  For the gas mass estimates, we
have assumed spherical clouds, which is probably a very poor
assumption.  As such, the estimated gas masses should be considered
upper limits.

Finally, the ionization modeling assumes photoionization and
collisional ionization equilibrium, which we have shown to be a valid
condition for the majority, but not all of the clouds.  In addition,
the metallicity estimates are based upon the assumption of single
phase ionization conditions.  If some of the plausible concerns
expressed above with regard to heterogeneous physical conditions
aligned in line of sight velocity hold, then multi-phase structure
could be present that would affect the metallicity estimates.
Overall, the assumption of ionization equilibrium in single phase gas
is critical to all deduced quantities, especially the metallicities
and the thermal equilibrium values presented in \S~\ref{sec:hydroeq}.

\section{Discussion}
\label{sec:discuss}

With a velocity extent of 1600~{\kms}, the {\HI} absorption complex at
$z=0.672$ in the quasar Q1317+277 is a most intriguing gaseous
structure.  Absorption with velocity spreads on the order of $2000$
{\kms} occur in approximately 10-15\% of quasars
\citep{weymann91,gibson09}. In almost all cases, extreme absorption of
this nature is produced by material ejected from the quasar itself
(i.e., broad absorption line quasars); however, the properties of the
$z=0.672$ {\HI} complex studied here are not suggestive of absorption
``associated'' with or ``intrinsic'' to the quasar.  For example, the
{\HI} exhibits no evidence of partial covering, which is an adopted
signature of associated gas \citep{barlow97,ganguly99}.  Furthermore,
the kinematics of the metals absorption lines are kinematically
similar to the velocity spreads observed in galaxy halos
\citep{cv01,cvc03}.  Thus, the {\HI} complex is likely inervening
absorption.  Early on, \citet{bs65} suggested that the environments of
galaxy clusters may give rise to extensive, intervening broad
absorption line complexes, but few potential candidates have been
identified.

In this section, we summarize and further examine the nature and
environment of the $z=0.672$ {\HI} complex toward Q1317+277, compare it
to other similar {\HI} complexes, and discuss the possible origin of
the Q1317+277 {\HI} complex, such as hot-mode or cold-mode accretion,
galactic winds, accreting filaments, intracluster gas, and/or the
warm-hot phase of the intergalactic medium (WHIM).

\subsection{The Nature of the $z=0.672$ {\HI} Complex}

The Q1317+277 {\HI} complex at $z=0.672$ is characterized by a velocity
spread of 1600~{\kms} and 21 {\Lya} components with $12.9 \leq \log
N({\HI}) \leq 14.5$, as determined by Voigt profile fitting.  Under
the assumption that the components (clouds) are near hydrodynamic
equilibrium, the temperatures, hydrogen number densities, and hydrogen
ionization fractions primarily range between $10^{4} \leq T \leq
10^{5}$~K, $ -3.9 \leq \log n_{\hbox{\tiny H}} \leq -4.9$, and $ -5.5
\leq \log f_{\hbox{\tiny HI}} \leq -6.0$.  The deduced cloud sizes are
on the order of 200 kpc, and the cloud baryonic gas masses range
between $10^9$--$10^{10}$~M$_{\odot}$.  Because the gas masses scale as
$n_{\hbox{\tiny H}}^{-2}f_{\hbox{\tiny HI}}^{-3}$, the large masses
result from the low hydrogen number densities and the high ionization
conditions of the clouds.

The metallicities are measured only for clouds \#7, \#8, and \#10 and
are $\log Z/Z_{\odot} = -1.7$, $-2.5$, and $-1.8$, respectively.
Upper limits on the remaining clouds are $\log Z/Z_{\odot} < -2.4$ to
$\log Z/Z_{\odot} < -1.7$.  The limits on the cloud metallicities do
not rule out enrichment at the level of the high redshift IGM
\citep[e.g.,][]{cowie98,simcoe04}.  On the other hand, the low
metallicities are 1--2 orders of magnitude below the $\log Z/Z_{\odot}
\sim -0.6$ metallicities measured in $z = 0.7$ X-ray clusters
\citep{balestra07,maughan08}.

Further insight is gained by examination of the kinematic-ionization
substructure.  Near the velocity center of the complex, at $v \simeq
75$~{\kms} with respect to the galaxy G1, is a hot $T=480,000$~K,
collisionally ionized component (cloud \#11), which is very likely
shocked gas.  The narrow velocity region just blueward of cloud \#11
comprises four clouds, within which the only metal lines are detected.
The general overall absorption morphology of these four components
(clouds \#7--10), is that of a double profile suggesting two absorbing
structures contiguous in velocity space.  Cloud \#10, separated by
$\simeq 50$~{\kms} from cloud \#11, has detected {\OVI} absorption,
which is deduced to arise in cool $T=20$--30~K photoionized gas.
Clouds \#7 and \#8, which have detected {\OVI}, {\CIV}, and {\CIII}
absorption, give rise to a single profile, which is best modeled with
a narrow core (cloud \#7, cool photoionized gas with $T \simeq
10,000$~K) and a broad component (cloud \#8, hot collisionally ionized
gas with $T=115,000$~K).  The velocity centroids are separated by less
than half of a single COS spectral resolution element of $\simeq
17$~{\kms}.  In addition, cloud \#9 appears to be a very narrow
($T<10,000$ K) blue wing of cloud \#10 offset by $\simeq 20$~{\kms}
that overlaps with the red wing of cloud \#8.  These substructures may
be suggestive of clouds moving through a hot, $T > 10^6$ K, medium in
which a conductive interfaces arises at the boundary between the cool,
warm, and hot gas \citep[e.g.,][]{sembach03}.


Knowing the environment of the {\HI} complex and relationship to
galaxies would be instrumental for a broader interpretation.  The
proximate galaxy is G1, which lies at $D=58$ kpc from the quasar line
of sight and has a redshift very near the mean of the {\HI} complex.
The virial mass of galaxy G1 is estimated to be within a factor of two
of the virial mass of M87 \citep[$M_{\rm vir}/M_{\odot} \simeq
10^{14}$,][]{strader11}, suggesting that galaxy G1 could be a central
galaxy in a Virgo-like cluster.  However, we find no clearly
compelling evidence that G1 resides in a galaxy cluster or in a group
with an X-ray emitting intracluster medium.

A search of the NASA Extragalactic Database (NED) and SIMBAD database
yielded no reported X-ray measurements of Q1317+277.  Within $4'$ of
Q1317+277, there are no sources in the ROSAT all-sky survey bright
source catalog \citep{mickaelian06,voges99}.  Of the five closest
X-ray sources that are not identified either as a star or an
AGN/quasar (which have known redshifts), one (1RXS J131954.5+253210)
lies at $2^{\circ}$ from Q1317+277 (50 Mpc projected at $z=0.672$).
The bright $R=15.1$ galaxy identified within 29~{\arcsec} of the
X-rays would be 2 Mpc projected from this source at $z=0.672$.  If the
X-ray source is associated with this galaxy, it is likely that the
galaxy and X-ray source reside at a redshift much lower than the {\HI}
complex.

In the ROSAT HR1 band, a minimum of 0.04 cnts s$^{-1}$ is required for
a source to be included in the ROSAT catalog. Applying this upper
limit, and invoking the relationship \citep{mullis01} between the
count rate and the total flux in the band (accounting for the aperture
correction), we estimate $L_{\hbox{\tiny X}} < 2 \times 10^{39}$ erg
s$^{-1}$.  This is four orders of magnitude below the expected X-ray
luminosity of $2 \times 10^{43}$ erg s$^{-1}$ for a cluster with a
central galaxy of virial mass of G1, where we have employed the
bolometric X-ray luminosity to virial mass scaling relation of
\citet{bryan98} and corrected for the X-ray band
\citep[see][]{mullis01}.  The virial temperature of galaxy G1 is
estimated to be on the order of $10^7$ K, which yields a coronal
temperature of $kT \simeq 0.8$ keV.  According to the compilation of
\citet{crain11}, our upper limit on $L_{\hbox{\tiny X}}$ is not
inconsistent with the observed X-ray luminosities of early-type
galaxies with similar $kT$.  

We have successfully measured spectroscopic redshifts for only
galaxies G1 and G2.  As such, we cannot directly deduce the presence
of a cluster at $z=0.672$, nor estimate the velocity dispersion of the
galaxies that may reside at this redshift.  
Based upon the photometric properties examined in the imaging date, it
is difficult to definitively rule out or favor the presence of a
cluster at $z=0.67$.
However, the upper limits on the X-ray flux within 50 Mpc projected
from Q1317+277 and the low metallicity of the {\HI} complex, 1-2 dex
below intracluster gas measurements \citep{balestra07,maughan08}, do
not favor a large cluster nor a hot intracluster medium.

Such considerations leave open the possibility that the {\HI} complex
may be a phenomenon closely linked to a massive old elliptical galaxy
that is not in an overdense environment.  Taken together at face
value, the data and the results of our analysis suggest a low
metallicity structure, possibly a filament or the remnants of a
disrupted filament.  There remains the question of the possible
connection to the smaller galaxy G3, which unfortunately does not have
a measured or estimated redshift.

\subsection{Review of Comparable {\HI} Complexes}

Given the dramatic velocity spread and kinematics of the {\HI} complex
toward Q1317+277, and given its proximity to galaxy G1 (and possibly
G3), it is of interest to investigate how rare/common are such
absorbing complexes, what their observed relationships are with
respect to galaxies, and what interpretations have been adopted in view
of the role of gas in the evolution of individual and group galaxies.
Such insights may help identify the Q1317+277 {\HI} complex in a
broader context.

One example is the $z \simeq 2$ {\CIV} absorber complex toward the
``Tololo Pair'' (Tol 1037--27, $z_{\rm em}=2.18$, and Tol 1038--27,
$z_{\rm em}= 2.33$), which may be produced by intracluster gas
\citep{jakobsen86}.  The {\CIV} doublets, later observed in two
additional quasars in proximity on the sky, exhibit multiple discrete
components with velocity widths ranging between 50--1000~{\kms} and
may extend some 18 Mpc \citep{dinshaw96}.

Another possible intracluster absorption complex, at $z=0.695$ toward
the $z_{em}=1.05$ quasar PG 2302+029 \citep{jannuzi96}, exhibits broad
($\Delta v = 3000$~{\kms}) high ionization {\CIV},
{\NV}, and {\OVI} doublets.  In the FOS spectrum (velocity resolution $\Delta v
=230$~{\kms}), the {\Lya} absorption is segregated into multiple
individual components each with $\Delta v < 250$~{\kms} distributed
across the full velocity range of the metals. Near the central
velocity, narrow {\CIV}, {\NV}, and {\OVI} are present in one {\Lya}
component.  No low ionization species are present.  \citet{jannuzi96}
suggest three possible interpretations: (1) material ejected from the
quasar at extreme ejection velocity, (2) material associated with
galaxies or the intracluster medium of a cluster or supercluster of
galaxies, and (3) remnant material from supernovae in a galaxy.
Unfortunately, their observations did not provide data capable of
distinguishing between these scenarios.

Toward the $z_{\rm em} = 0.297$ quasar H1821+643, \citet{tripp01}
reported an {\HI} complex at $z=0.1212$ comprising five {\Lya}
components distributed over a velocity interval of $\sim 700$~{\kms}
in high resolution STIS and {\it FUSE\/} spectra.  The $\log N({\HI})$
column densities range from 12.7--13.8.  Absorption from {\OVI} is
present in a single broad wing of the central component, for which
collisional ionization is favored with $T = 10^{5.3-5.6}$ K and $-1.8
\leq \hbox{[O/H]}\leq -0.6$.  Seven galaxies in the velocity range of
the absorption are present at impact parameters ranging from 140--2400
kpc, with the 140 kpc galaxy aligned in redshift with the {\OVI}
absorption. \citet{tripp01} favor the scenario in which the {\HI}
complex is intragroup gas or an unvirialized filamentary structure.

Using GHRS, STIS and {\it FUSE\/} spectra of the $z_{\rm em} = 0.116$
BL Lac object PKS 2155--304, \citet{shull98} and \citet{shull03}
analyzed an {\HI} complex with 14 {\Lya} components centered at
$z=0.056$ spread over a velocity interval of 2270~{\kms}.  They
estimate the {\HI} column densities have the range $14.5 \leq \log
N({\HI}) \leq 15.0$ with $\log Z/Z_{\odot} < -2.5$ and cloud depths
less than 400 kpc.  Five {\HI} emitting galaxies are found in the
range $0.056 \leq z \leq 0.058$ with impact parameters 400--790
kpc. The two strongest {\Lya} blends have detectable {\OVI} and
possible {\OVIII} absorption measured in a {\it Chandra\/} spectrum.
If the gas is the warm-hot ionized medium (WHIM) then the density is
constrained to $\log n_{\hbox{\tiny H}} \simeq -4$.  \citet{shull03}
favor a scenario in which the {\OVI} arises in ``nearside'' and
``backside'' shocked infall into the potential well of the galaxy
group.


In high resolution STIS and {\it FUSE\/} spectra of the $z_{\rm em} =
0.370$ quasar HS 0624+6907, \citet{aracil06a} report a cluster of 13
{\Lya} lines\footnote{An ``erratum'' to \citet{aracil06a} was
  published \citep[see][]{aracil06b}.  However, the deduced properties
  of the {\HI} absorbing complex are unaltered.} at $z=0.0635$ with a
velocity spread of 1000~{\kms}.  The {\HI} column densities range
between $12.6 \leq \log N({\HI}) \leq 15.3$.  Only in the central
component, with total $\log N({\HI}) = 15.4$, are metal lines detected
({\SiIII}, {\SiIV}, and {\CIV}, but no {\OVI}) from which the gas is
deduced to be photoionized with metallicity $\log Z/Z_{\odot} =
-0.05$, very near to solar enrichment, with $\log n_{\hbox{\tiny H}} =
-3.9$.  The gas temperatures are deduced to be $T < 10^{5}$~K.  The
estimated baryonic mass of this component is $\sim 10^5$~M$_{\odot}$
with an absorption length scale of 3--5 kpc.  They report 10 galaxies
within 135--1370 kpc in the range $0.062 \leq z \leq 0.067$, but this
group is not consistent with elliptical-rich groups.  On account of
the high metallicity and cool temperatures, \citet{aracil06a} favor
the interpretation that this {\HI} complex is tidally stripped
material from one of the nearby galaxies.

The {\HI} complexes toward H1821+643, PKS 2155--304, and HS 0624+6907
have both similarities and differences with the {\HI} complex toward
Q1317+277.  However, the Q1317+277 {\HI} complex bares little
resemblance to the metal-rich complexes observed toward PG 2303+029
and toward the Tol 1037--27 and Tol 1038--27 pair.  These latter two
complexes may be examples of metal enriched intracluster gas.

The broad {\HI} component in the complex toward H1821+643 exhibits
{\OVI} that is likely to be predominantly collisionally ionized with a
relatively high metallicity.  Similarly, cloud \#8 in the Q1317+277
{\HI} complex appears to be a $T > 10^5$ K, collisionally ionized
{\OVI} absorber, but accompanied by {\CIV} and {\CIII} absorption.  On
the other hand, the hottest, broad component in the Q1317+277 {\HI}
complex has no detected {\OVI} and has upper limits on metallicity
indicating that it is metal poor in comparison.  The H1821+643 {\OVI}
absorber is at a substantially larger impact parameter to the nearest
galaxy, which resides in a group that clearly has no massive
elliptical galaxy, whereas the Q1317+277 {\HI} complex quite is very
close in projected to the massive elliptical galaxy G1.

The {\HI} column density for the low metallicity {\OVI} absorber in
the complex absorption toward PKS 2155--304 is 1--2 orders of
magnitude greater than the $N({\HI})$ of the components in the
Q1317+277 {\HI} complex.  However, the clouds have similar
$n_{\hbox{\tiny H}}$.  In both complexes, the {\OVI} resides to the
wings of largest {\Lya} components.  \citet{shull03} interpret this as
a shock interface, and this interpretation may apply in the case of
the Q1317+277 {\HI} complex.  However, as with the H1821+643 {\OVI}
absorber, the environment of the PKS 2155--304 {\HI} complex resides
within a moderate group of galaxies having no massive elliptical
galaxy.

Of the three examples, the {\HI} complex toward HS 0624+6907 has an
{\HI} absorption profile morphology most similar to that of the
Q1317+277 {\HI} complex.  The cool photoionized cloud with {\CIV},
{\SiIV}, and {\SiIII} absorption compares to the cool photoionized
{\OVI} absorbing cloud \# 10 in the Q1317+277 {\HI} complex, but cloud
\# 10 has no {\SiIV} or {\CIV} absorption.  Though the clouds have
similar $n_{\hbox{\tiny H}}$, cloud \#10 has lower $N({\HI})$, higher
ionization conditions, and a much lower metallicity as compared to the
nearly solar metallicity cloud in the HS 0624+6907 complex.  Because
of the low ionization deduced for the latter cloud, the mass is $\sim
4$ orders of magnitude smaller than the mass deduced for cloud \# 10.
Again, the galaxy environment of the HS 0624+6907 {\HI} complex
contains no massive elliptical galaxy.

Overall, two unique features to the Q1317+277 {\HI} complex are that
it is at substantially higher redshift ($z=0.672$) compared to the
other reported {\HI} complexes ($z = 0.056$, $0.064$ and $0.121$)
and that it is in close projected proximity to a region dominated by
massive, metal-rich elliptical with an old stellar population.

\subsection{Interpreting the $z=0.672$ {\HI} Complex}

Summarizing the gas properties of the Q1317+277 {\HI} complex, we find
(1) a hot, photo and collisionally ionized component that is
consistent with shocked gas, (2) a cool component with photoionized
{\OVI} absorption, and (3) a cool component plausibly layered within a
warm component that is both photo and collisionally ionized and
exhibits {\OVI}, {\CIV} and {\CIII} absorption, (4) several additional
warm {\HI} components spread over 1600~{\kms} in the rest-frame of
$z=0.672$, and (5) measurements of and limits on metal-enrichment
between $-2.5 \leq \log Z/Z_{\odot} \leq -1.7$.

If the rest-frame velocity spread of the {\HI} complex is due to the
local Hubble flow, then the line-of-sight proper length of the
structure would be $D_{\hbox{\tiny HF}} = \Delta v/H_0E(z)$, where
$\Delta v = 1600$ {\kms} and $E^2(z) =
\Omega_m(1+z)^3+\Omega_\Lambda$.  At $z=0.67$, we estimate
$D_{\hbox{\tiny HF}} \sim 15$~Mpc.  Based upon the ionization
modeling, the deduced physical sizes of the the {\HI} complex are not
consistent with a single structure of this extent.  If the velocities
of the {\HI} absorbers are due to Hubble flow, then the complex must
comprise absorbers that are spatially segregated; there are five main
absorption features apparent in the {\Lya} profile, which would imply
an average separation of 3 Mpc and that we have by chanced probed
several isolated absorption systems.  Considering the similarities of
the properties of the {\HI} complexes toward H1821+643, PKS 2155--304,
and HS 0624+6907, and that they are clearly connected with galaxies on
scales of 0.1--1 Mpc, we do not favor the interpretation that the
Q1317+277 {\HI} complex is multiple individual absorbers tracing a 15
Mpc Hubble flow.

Is it reasonable that the individual peaks in the {\HI} column density
are a result of having fragmented from a single structure?  If this
structure were Jean's unstable, i.e., $\tau_{\rm dyn} \ll \tau_{\rm
sc}$, then fragmentation and/or shock disruption into the observed
components would be plausible.

Crudely modeling this hypothetical single structure as having the
$N({\HI})$ weighted mean temperature of the components, $\left< \log T
\right> = 4.65$ (prior to fragmentation or shock disruption, where the
latter could subsequently heat the shocked components), and assuming
that the length of its long axis is the sum of the Jean's lengths for
each of the components, we find that $\tau_{\rm dyn} \simeq
\sfrac{1}{20} \tau_{\rm sc}$ independent of the assumed mean hydrogen
density, $\bar{n}_{\hbox{\tiny H}}$, where $\tau_{\rm dyn} \simeq
3\times 10^9 \, ( \bar{n}_{\hbox{\tiny H}} /10^{-4})^{1/2}$ and
$\tau_{\rm sc} \simeq 6\times 10^{10} \, (\bar{n}_{\hbox{\tiny
H}}/10^{-4})^{1/2}$ yr.  The model is consistent with the plausibility
that the multiple component structure of the {\HI} absorption complex
could very well have resulted from the fragmentation of a
quasi-coherent single gaseous structure.

Our simple model yields a total Jean's length of $L_J \simeq 2 \,
(\bar{n}_{\hbox{\tiny H}}/10^{-4})^{1/2}$ Mpc and a total Jean's mass
of $M_J \simeq 2 \times 10^{11} \, (\bar{n}_{\hbox{\tiny
H}}/10^{-4})^{1/2}$ Mpc, where the total mass is simply the sum of the
Jean's masses for the individual components.  The implied aspect ratio
$\beta = 2R/L$ for a cylindrical structure is $\beta \simeq (M_J/
\bar{\rho}_g L_J^3 ) \simeq 0.1$.  If $\log \bar{n}_{\hbox{\tiny H}} =
-4$, then the {\HI} complex could be crudely envisioned as a
$10^{11}$~M$_{\odot}$ mass cylindrical filament 2 Mpc long with a 200
kpc diameter.  These values would scale as $(\bar{n}_{\hbox{\tiny
H}})^{1/2}$.

Summarizing the luminous environment of the {\HI} complex, we find (1)
a massive, red, high-metallicity, elliptical galaxy (G1) with an old
stellar population ($\sim 6$ Gyr) residing at $D=58$ kpc and aligned
in velocity {\it between\/} the shocked {\HI} component and the
metal-enriched {\HI} components, near the {\HI} profile velocity
centroid, and (2) the upper limits on the X-ray luminosity 
consistent with the measured range for ellipticals with $kT \simeq
0.8$ keV and well below expected values for clusters dominated by
massive ellipticals.

In addition to the X-ray data not supporting the idea that galaxy G1
resides in a large cluster, the metallicity of the {\HI} absorbing gas
is constrained to be 1-2 orders of magnitude below the average
enrichment of clusters at $z=0.7$ \citep{balestra07,maughan08}.  The
observations of \citet{brinchman00} and \citet{vanderwel05} indicate
that old, massive, metal-rich late-type galaxies were present in the
field as early as $z=2$--$3$.  The stellar population models for
galaxy G1 suggest a formation epoch at $z\simeq 4$.  It may be that
galaxy G1 is not a dominant elliptical galaxy of a large group, but
may be a galaxy in the field that formed at high redshift.

In addition to the aforementioned supporting observations, a massive,
red, metal-rich elliptical far from a cluster environment is
theoretically plausible. The cosmological simulations of
\citet{gabor12}, which treat accretion and feedback processes, yield a
substantial fraction of red, high mass galaxies independent of
overdensity.  \citet{johansson12} demonstrate that massive early-type
galaxies are built in two phases.  The first is an initial growth via
in situ star formation fed by cold accretion ($z > 3$) that is later
quenched via virial shocking
\citep[e.g.,][]{birnboim03,keres05,dekel06,keres09,vandevoort11,vandevoort+scahye11}.
The second phase is accretion of stellar material ($z \sim
1.5$--$0.5$), called ``dry minor mergers'' due to the fact that their
gas is heated and stripped by shock heating in the hot virial halo of
the massive galaxy \citep[e.g.,][]{khochfar09,hopkins10}.  These
general results are also found by \citet{naab07,naab09}.

With a mass of $\log M_{\rm vir}/M_{\odot} > 13$, galaxy G1 is well
above the ``critical mass'' where the cooling time of the gas is much
longer than the dynamical time, such that an accretion shock is
established near the virial radius
\citep[e.g.,][]{birnboim03,dekel06,keres09,vandevoort+scahye11}.  The
rate at which the cooling time increases with decreasing redshift
[$\tau_{\rm cool} \propto 1/\rho \propto (1+z)^{-3}$] is more rapid
than the increase in the dynamical time with decreasing redshift,
[$\tau_{\rm dyn} \propto 1/\sqrt{\rho} \propto (1+z)^{-3/2}$].  Thus,
it is generally found in simulations that, at high redshifts, cold
streams can often penetrate the hot halos and accrete onto the
galaxies, whereas at intermediate to low redshifts, the longer cooling
time results in the heating and shocking of the cold filaments, which
then accrete into the halo but not onto the galaxy
\citep{keres05,dekel06,faucher-gigure11,vandevoort11}. Galaxy G1 may
have experienced the process in which it formed stars early during its
initial accretion, perhaps even experiencing cold accretion from the
filament that we are observing, and then at later times the accretion
penetrated no further than into the halo, resulting in a high mass,
metal-rich, early-type galaxy with quenched star formation.

Given such a scenario, the {\HI} complex could be interpreted as a
filamentary structure with IGM chemical enrichment levels that is
undergoing shock disruption near the viral radius of galaxy G1.  The
hot cloud (\#11) exhibits the signature of the shocked portion of the
filament, and the multiphase absorbers (clouds \#8-10) exhibit the
signature of a conductive interface.  Galaxy G1 is probed by the
quasar sight line impact parameter at $D/R_{\rm vir} \simeq 0.1$.
Simulations indicate that ``cold'' filaments (i.e., those that have
not been heated above $\log T = 5.5$) penetrate no deeper than $R
\simeq 0.5R_{\rm vir}$ at intermediate redshifts
\citep{keres09,vandevoort11}, suggesting that the sight line is
intercepting the filament in the outer part of the virialized halo.

\begin{figure*}[thb]
\plotone{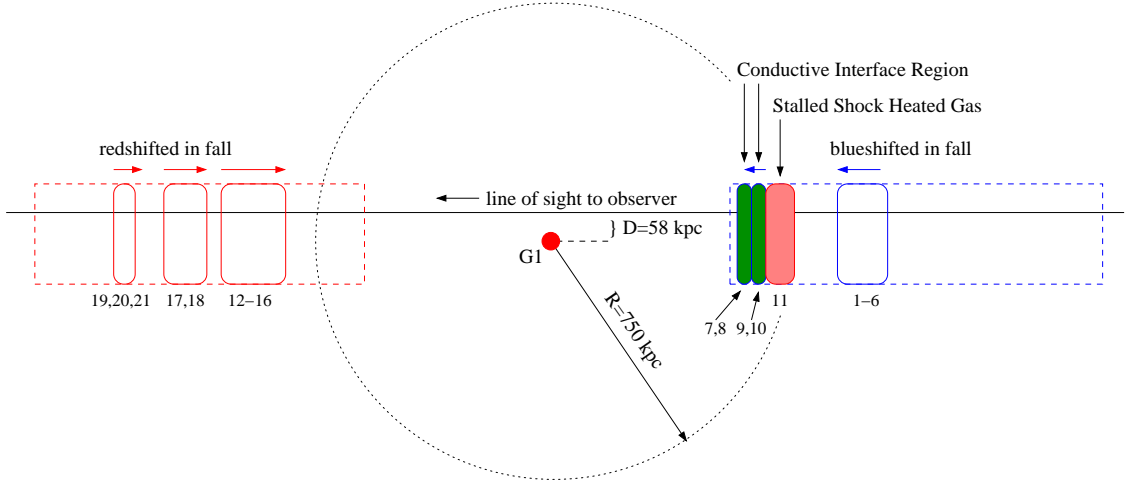}
\caption{A cartoon model illustrating the interpretation of the {\HI}
absorbing complex as an accreting filament onto the galaxy G1.  The
positive velocity material would require that it is infalling from the
oberver's side of the galaxy, whereas the negative velocity material
would be infalling opposite the observer's side of the galaxy.  The
plausible relative physical locations of the absorbing clouds, labeled
by their VP component numbers (see Table~\ref{tab:hiresult}), are
shown.  The conductive interface (clouds \#7-10) continue their infall
after being stalled in a shock front, which is presently coincident
with cloud \#11 (see text).  The model is adapted from simulation
results \citep[e.g., see the ``cold only'' panel of Figure 6 in][]{keres09}.}
\label{fig:cartoon}
\end{figure*}

A cartoon model of this interpretation is illustrated in
Figure~\ref{fig:cartoon}.  The cartoon is consistent with simulation
results and is guided by inspection of Figure 7 (top panel, for a
$z=2$, $M_{\rm vir} = 10^{12}$~M$_{\odot}$ galaxy) from
\citet{vandevoort11} and the ``cold only'' panel of Figure 6 for a
$z=1$ , $M_{\rm vir} = 10^{13}$~M$_{\odot}$ galaxy from
\citet{keres09}.

Presumably, any components of the filament that have not yet been
shocked will be shock heated and their cooling time will increase to
be on the order of the Hubble time (as is deduced for cloud \#11).
Those components that may reside within the shock radius will likely
not survive past $R \simeq 0.5R_{\rm vir}$ and be assimilated into the
hot halo.  This would imply that, ultimately, the filament will
accrete into the halo, be heated to the halo temperature, and never
accrete onto the galaxy itself.  Galaxy G1 has likely not had new gas
to fuel star formation for several Gigayears and will likely not
acquire new gas via accretion as it evolves.  Apart from the
possibility of dry minor mergers
\citep{khochfar09,hopkins10,johansson12} building the stellar mass of
galaxy G1, it is likely that the galaxy has experienced secular
stellar evolution since $z=4$.

If galaxy G3 is at $z \simeq 0.67$, one might ask if the filament is
associated with this lower mass $\log M_{\rm vir}/M_{\odot} \simeq
11.9$ galaxy.  With the data in hand, it is nearly impossible to claim
any relationship between galaxy G3, the {\HI} complex, and galaxy G1.
However, a virial shock is not predicted for lower mass galaxies,
which have cooling times shorter than their dynamical times.  As such,
it would be expected that a filament accreting onto galaxy G3 would
not show the signature of shocked gas with conductive interfaces as
seen in the {\HI} complex.  It could be speculated that galaxy G3 is
embedded in the filament, perhaps contributing to some in situ metal
enrichment, and might eventually be a dry minor merger (2\% mass
ratio) with galaxy G1.

This scenario for the evolution of galaxy G1 is in stark contrast to
the evolution of spiral galaxy G2, which is observed to have on-going
star formation at $z=0.661$, likely due to on-going multiphase
accretion that is feeding the galaxy \citep{ggk-q1317}.  To the limit
of $\log N({\HI}) \simeq 12.4$, there is no indication in the COS
spectrum of absorbing gas that bridges these two galaxies, which are
separated by $\Delta v = 1960$~{\kms} (in the frame of the galaxies).

The scenario of a shock heated filament for which the accretion onto
galaxy G1 is quenched is very different than the favored scenarios for
the lower redshift {\HI} complexes toward H1821+643, PKS 2155--304,
and HS 0624+6907 \citep{shull98,shull03,tripp01,aracil06a}.  None of
these complexes are in the vicinity of a massive elliptical galaxy,
and as such it would be expected that they represent physical
scenarios other than the one we propose for the Q1317+277 {\HI}
complex.  That is, {\HI} absorbing complexes, though filamentary in
nature, likely trace various environments around galaxies and their
relationships with the intergalactic medium.


\section{Conclusions}
\label{sec:conclude}

We have studied the gas properties of and the luminous environment
around a remarkable {\HI} absorbing complex over the range
$z=0.6686$--$0.6768$ having a velocity spread of $\Delta v =
1600$~{\kms} towards the quasar Q1317+277.  To constrain the gas
absorption properties, we have analyzed COS, STIS, and HIRES spectra
of the quasar to examine the Lyman series, {\MgII}, {\CIV}, and {\OVI}
absorption using Voigt profile fitting and photo+collisional
ionization modeling.  The galaxy properties in the quasar field have
been measured using a WFPC2/F702W image, and multi-band APO/SPIcam and
KPNO/IRIM ground-based images.  Stellar population modeling and halo
abundance matching were employed to estimate the stellar ages,
metallicities, mass, and virial mass, radii, and temperatures of the
galaxies.

{\it Absorption Properties:} 
The {\HI} complex is characterized by five contiguous absorbing
regions, comprising 21 subcomponents, or clouds.  Ionization models
suggest that the kinematics are not due to the local Hubble flow, but
are consistent with a $\sim 2$ Mpc by $\sim 200$ kpc structure with a
total gas mass on the order of $10^{11}$~M$_{\odot}$.  We find a hot
$\log T = 5.7$ cloud, which we attribute to shock heated gas.  This
cloud is bordered by multiphase, cool, warm, and hot layers gas
suggestive of a conductive interface in which metal line absorption
({\CIII}, {\CIV}, and {\OVI}) is detected.  The low metallicity of the
gas ($-2.5 \leq \log Z/Z_{\odot} \leq -1.7$) is more consistent with
enrichment levels of the high redshift intergalactic medium
\citep[e.g.,][]{cowie98,simcoe04}, perhaps with low levels of in situ
enrichment, than with the metallicities observed in the intracluster
medium at $z=0.7$ \citep{balestra07,maughan08}.

{\it Galaxy Properties:}
We find that the galaxy G1 ($z=0.6719$ at impact parameter $D=58$ kpc)
is consistent with a massive ($\log M_{\rm vir}/M_{\odot} \simeq
13.7$) elliptical galaxy that is metal-rich ($Z \simeq Z_{\odot}$), and
formed at $z=4$ (6 Gyr stellar population).  The magnitudes and colors
of the other galaxies in the field, limits on the X-ray luminosity,
and the low metallicity of the {\HI} absorbing gas are consistent with
galaxy G1 being a field galaxy with an evolutionary scenario in which
the cold accretion has been shut down by shock heating.  The redshift
of galaxy G1 ($z=0.6719$) is bracketed by the redshifts of the
collisionally ionized gas ($z=0.6723$, $\Delta v = +76$~{\kms}) and
the multiphase gas with detected metals consistent with a conductive
interface ($z=0.6712$--$0.6717$, $\Delta v = -124$ to $-30$~{\kms}) in
the {\HI} complex.

{\it Interpretation:}
Based upon our data and analysis, we favor the scenario in which the
{\HI} complex is a filamentary structure accreting into the halo of
galaxy G1 that is experiencing virial shock heating and dynamical
disruption.  Consistent with predictions of both theoretical
treatment \citep[e.g.,][]{birnboim03,dekel06,birnboim07} and
cosmological simulations
\citep[e.g.,][]{keres05,keres09,khochfar09,hopkins10,faucher-gigure11,vandevoort11,vandevoort+scahye11,johansson12}
of massive galaxies (i.e., $\log M_{\rm vir}/M_{\odot} > 12$), our
observations and analysis indicate that the gas accreting into the
halo of galaxy G1 at times after its formation epoch have not accreted
onto the galaxy itself and that the {\HI} complex will also not
accrete onto the galaxy.  As such, the star formation of galaxy G1 has
likely been quenched for Gigayears.

Though it is difficult to definitively determine the nature of the
{\HI} absorbing complex, the scenario we favor is highly plausible and
consistent with simulations and theory.  In fact, the data appear to
provide convincing evidence that theory and simulations correctly
predict by $z<1$, cold accretion via filaments in high mass galaxies
is shock heated as it accreted into the virialized hot halos and that
this gas then grows the halos, but does not fuel further star
formation.  Continued growth of the stellar mass would then
necessarily occur via dry mergers of minor galaxies.

\acknowledgments

We thank Daniel Ceverino for several stimulating and informative
discussions during his visit to New Mexico State University and for
helpful comments on an early draft of this paper.  We also thank Kyle
Stewart for several informative email exchanges with regard to halo
abundance matching methods.  This research was primarily support
through grant HST-GO-11667.01-A provided by NASA via the Space
Telescope Science Institute, which is operated by the Association of
Universities for Research in Astronomy (AURA) under NASA contract NAS
5-26555.  CWC thanks GGK, and Michael T. Murphy, and Swinburne Faculty
Research Grants for providing funding for a visit to Swinburne
University of Technology. Some observations are obtained with the
Apache Point Observatory 3.5-meter telescope, which is owned and
operated by the Astrophysical Research Consortium (ARC).  Additional
data were obtained at Kitt Peak National Optical Astronomy
Observatory, which is operated by AURA under cooperative agreement
with the National Science Foundation.  Some data presented herein were
obtained at the W.M. Keck Observatory, which is operated as a
scientific partnership among the California Institute of Technology,
the University of California and NASA.  The Observatory was made
possible by the generous financial support of the W.~M. Keck
Foundation. The authors wish to recognize and acknowledge the very
significant cultural role and reverence that the summit of Mauna Kea
has always had within the indigenous Hawaiian community.  This
research has made use of the NASA/IPAC Extragalactic Database (NED)
which is operated by the Jet Propulsion Laboratory, California
Institute of Technology, under contract with NASA.  This research has
also made use of the SIMBAD database, operated at Centre de
Donn\'{e}es, Strasbourg, France.

{\it Facilities:} \facility{HST (WFPC2, STIS, COS)}, \facility{Keck
I (HIRES, LRIS)}, \facility{APO (SPIcam)}, \facility{KPNO (IRIM)}.


\begin{thebibliography}{99}

\bibitem[Abraham {\etal}(1996)]{abraham96} Abraham, R.~G., van den
Bergh, S., Glazebrook, K., Ellis, R.~S., Santiago, B.~X., Surma, P.,
\& Griffiths, R.~E.\ 1996, \apjs, 107, 1

\bibitem[Aracil {\etal}(2006a)]{aracil06a} Aracil, B., Tripp, 
T.~M., Bowen, D.~V., {\etal} 2006a, \mnras, 367, 139

\bibitem[Aracil {\etal}(2006b)]{aracil06b} Aracil, B., Tripp, 
T.~M., Bowen, D.~V., {\etal} 2006b, \mnras, 372, 959

\bibitem[Asplund {\etal}(2009)]{asplund09} Asplund, M., Grevesse, N.,
Sauval, A.~J., \& Scott, P.\ 2009, \araa, 47, 481

\bibitem[Bahcall {\etal}(1993)]{cat1} Bahcall, J.~N., {\etal} 1993,
ApJS, 87, 1

\bibitem[Bahcall {\etal}(1996)]{cat2} Bahcall, J.~N., Bergeron, J.,
Boksenberg, A., {\etal} 1996, \apj, 457, 19

\bibitem[Bahcall \& Salpeter(1965)]{bs65} Bahcall,
J.~N., \& Salpeter, E.~E.\ 1965, \apj, 142, 1677

\bibitem[Balestra {\etal}(2007)]{balestra07} Balestra, I.,
Tozzi, P., Ettori, S., {\etal} 2007, \aap, 462, 429

\bibitem[Barlow \& Sargent(1997)]{barlow97} Barlow, T.~A.,
\& Sargent, W.~L.~W.\ 1997, \aj, 113, 136


\bibitem[Behroozi, Conroy, \& Wechsler(2010)Behroozi
{\etal}]{behroozi10} Behroozi, P.~S., Conroy, C., \& Wechsler, R.~H.\
2010, \apj, 717, 379

\bibitem[Bell {\etal}(2003)]{bell03} Bell, E.~F., McIntosh, D.~H.,
Katz, N., \& Weinberg, M.~D.\ 2003, \apjs, 149, 289

\bibitem[Bertin \& Arnouts(1996)]{bertin96} Bertin, E., \& Arnouts,
S.\ 1996, \aaps, 117, 393

\bibitem[Binney(1977)]{binney77} Binney, J.\ 1977, \apj, 215, 483

\bibitem[Birnboim \& Dekel(2003)]{birnboim03} Birnboim, Y., \& Dekel,
A.\ 2003, \mnras, 345, 349

\bibitem[Birnboim {\etal}(2007)]{birnboim07} Birnboim, Y., Dekel, 
A., \& Neistein, E.\ 2007, \mnras, 380, 339 

\bibitem[Bostroem {\etal}(2010)]{stis-ihb} Bostroem, K., A., {\etal}
2010, STIS Instrument Handbook, 2010-10.0, (Baltimore: STScI).

\bibitem[Brinchmann \& Ellis(2000)]{brinchman00} Brinchmann, J., \&
Ellis, R.~S.\ 2000, \apjl, 536, L77


\bibitem[Bruzual \& Charlot(2003)]{bruzual03} Bruzual, G., \& Charlot,
S.\ 2003, \mnras, 344, 1000

\bibitem[Bryan \& Norman(1998)]{bryan98} Bryan, G. L., \& Norman,
M. L. 1998, ApJ, 495, 80

\bibitem[Chabrier(2003)]{chabrier03} Chabrier, G.\ 2003, \pasp, 115,
763

\bibitem[Churchill(1997)]{cwc-thesis} Churchill, C. W. 1997,
Ph.D. Dissertation, University of Califormia, Santa Cruz

\bibitem[Churchill \& Klimek(2012)]{cwc-rates} Churchill, C. W., \&
Klimek, E.  2012, ApJ, in prep

\bibitem[Churchill {\etal}(2012)]{cwc-anatomy} Churchill, C.~W., Kacprzak,
G.~G., Nielsen, N. M., Steidel, C.~C., \& Murphy, M.~T.\ 2012, \apj, submitted

\bibitem[Churchill {\etal}(2007)]{paper1} Churchill, C.~W., Kacprzak,
G.~G., Steidel, C.~C., \& Evans, J.~L.\ 2007, \apj, 661, 714 (Paper I)

\bibitem[Churchill {\etal}(2000)]{archiveI} Churchill, C.~W., Mellon,
R.~R., Charlton, J.~C., {\etal} 2000, \apjs, 130, 91

\bibitem[Churchill {\etal}(1999b)]{weakI}
Churchill, C. W., Rigby, J. R., Charlton, J. C., \& Vogt, S. S. 1999b,
ApJS, 120, 51

\bibitem[Churchill \& Vogt(2001)]{cv01}
Churchill, C. W., \& Vogt, S. S. 2001, AJ, 122, 679

\bibitem[Churchill {\etal}(2003)Churchill, Vogt, \& Charlton]{cvc03}
Churchill, C. W., Vogt, S. S., \& Charlton, J. C. 2003, AJ, 125, 98


\bibitem[Conroy \& Wechsler(2009)]{conroy09} Conroy, C., \& Wechsler,
R.~H.\ 2009, \apj, 696, 620

\bibitem[Cowie {\etal}(1996)]{cowie96} Cowie, L.~L., Songaila, A., Hu,
E.~M., \& Cohen, J.~G.\ 1996, \aj, 112, 839

\bibitem[Cowie \& Songaila(1998)]{cowie98} Cowie, L.~L., \& Songaila,
A.\ 1998, \nat, 394, 44

\bibitem[Crain {\etal}(2011)]{crain11} Crain, R. A., McCarthy, I. G.,
Schaye, J., Frenk, C. S. \& Thuens, T. 2011, MNRAS, arXiv:1011.1906

\bibitem[Dav{\'e} {\etal}(1999)]{dave99} Dav{\'e}, R., 
Hernquist, L., Katz, N., \& Weinberg, D.~H.\ 1999, \apj, 511, 521 

\bibitem[Dekel \& Birnboim(2006)]{dekel06} Dekel, A., \&
Birnboim, Y.\ 2006, \mnras, 368, 2

\bibitem[Ding {\etal}(2005)Ding, Charlton, \& Churchill]{ding05} Ding, J.,
Charlton, J.~C., \& Churchill, C.~W.\ 2005, \apj, 621, 615

\bibitem[Dinshaw \& Impey(1996)]{dinshaw96} Dinshaw, N., \& Impey,
C.~D.\ 1996, \apj, 458, 73

\bibitem[Dixon {\etal}(2010)]{cos-ihb}
Dixon, W. V., {\etal} 2010, Cosmic Origins Spectrograph Instrument
Handbook, Version 3.0 (Baltimore: STScI)

\bibitem[Dopita \& Sutherland(2003)]{diffuseuniverse} Dopita, M. A.,
\& Sutherland, R. S. 2003, Astrophysics of the Diffuse Universe,
Springer

\bibitem[Draine(2011)]{draine11} Draine, B.~T.\ 2011, Physics of the
Interstellar and Intergalactic Medium, Princeton University Press,
ISBN: 978-0-691-12214-4 

\bibitem[Erb {\etal}(2006)]{erb06} Erb, D.~K., Steidel, C.~C.,
Shapley, A.~E., Pettini, M., Reddy, N.~A., \& Adellberger, K.~L.\
2006, \apj, 646, 107

\bibitem[Faber {\etal}(2007)]{faber07} Faber, S. M., {\etal} 2007,
ApJ, 665, 265

\bibitem[Ferland {\etal}(1998)]{ferland98} Ferland, G.~J., 
Korista, K.~T., Verner, D.~A., {\etal} 1998, \pasp, 110, 761 

\bibitem[Fontana {\etal}(2004)]{fontana04} Fontana, A., Pozzetti, L.,
Donnarumma, I., {\etal}\ 2004, \aap, 424, 23

\bibitem[Faucher-Gigu{\`e}re et al.(2011)]{faucher-gigure11}
Faucher-Gigu{\`e}re, C.-A., Kere{\v s}, D., \& Ma, C.-P.\ 2011,
\mnras, 417, 2982

\bibitem[Gabor \& Dav{\'e}(2012)]{gabor12} Gabor, J.~M., \& Dav{\'e},
R.\ 2012, arXiv:1202.5315

\bibitem[Ganguly {\etal}(1999)]{ganguly99} Ganguly, R., Eracleous, 
M., Charlton, J.~C., \& Churchill, C.~W.\ 1999, \aj, 117, 2594

\bibitem[Gibson {\etal}(2009)]{gibson09} Gibson, R.~R., Jiang, 
L., Brandt, W.~N., {\etal} 2009, \apj, 692, 758 

\bibitem[Hanson(1986)]{dqed} Hanson, R. 19986, Least Squares with
Bounds and Linear Constraints, SIAM Journal of Scientific and
Statistical Computing, Vol.\ 7 No.\ 3, 826

\bibitem[Hopkins et al.(2010)]{hopkins10} Hopkins, P.~F., Bundy, 
K., Hernquist, L., Wuyts, S., \& Cox, T.~J.\ 2010, \mnras, 401, 1099 

\bibitem[Johansson {\etal}(2012)]{johansson12} Johansson, P.~H., Naab,
T., \& Ostriker, J.~P.\ 2012, arXiv:1202.3441

\bibitem[Jakobsen {\etal}(1986)]{jakobsen86} Jakobsen, P., 
Perryman, M.~A.~C., di Serego Alighieri, S., Ulrich, M.~H., 
\& Macchetto, F.\ 1986, \apjl, 303, L27 

\bibitem[Jannuzi {\etal}(1996)]{jannuzi96} Jannuzi, B.~T., Hartig,
G.~F., Kirhakos, S., {\etal} 1996, \apjl, 470, L11

\bibitem[Jarosik {\etal}(2011)]{wmap} Jarosik, N., {\etal} 2011, ApJS,
192, 14

\bibitem[Kacprzak {\etal}(2011)]{kcems} Kacprzak, G. G., Churchill,
C. W., Evans, J. L., Murphy, M. T., \& Steidel, C. C 2011, MNRAS,
416, 3118 

\bibitem[Kacprzak {\etal}(2012)]{ggk-q1317} Kacprzak, G. G.,
Churchill, C. W., Steidel, C. C., Spitler, L. R, Holtzman, J. A., \&
Bouch\'e, N. A. 2012, MNRAS, submitted

\bibitem[Kannappan(2004)]{kannappan04} Kannappan, S.~J.\ 2004, 
\apjl, 611, L89 

\bibitem[Kere{\v s} {\etal}(2009)]{keres09} Kere{\v s}, D., Katz, N.,
Fardal, M., Dav{\'e}, R., \& Weinberg, D.~H.\ 2009, \mnras, 395, 160

\bibitem[Kere{\v s} {\etal}(2005)]{keres05} Kere{\v s}, D., Katz, N.,
Weinberg, D.~H., \& Dav{\'e}, R.\ 2005, \mnras, 363, 2

\bibitem[Khochfar \& Silk(2009)]{khochfar09} Khochfar, S., \& Silk,
J.\ 2009, \mnras, 397, 506

\bibitem[Kriss(2011)]{kriss11} Kriss, G. A., COS Instrument Handbook 2011-01,
(Baltimore, STScI)

\bibitem[Lawton {\etal}(2008)]{lawton08} Lawton, B., Churchill, C.~W.,
York, B.~A., {\etal} 2008, \aj, 136, 994

\bibitem[Lequeux(2005)]{lequeux} Lequeux, J. 2005, The Interstellar
Medium, Springer, ISBN: 3-540-21362-0 


\bibitem[Maughan {\etal}(2008)]{maughan08} Maughan, B.~J., Jones, 
C., Forman, W., \& Van Speybroeck, L.\ 2008, \apjs, 174, 117 

\bibitem[McGaugh(2005)]{mcgaugh05} McGaugh, S.~S.\ 2005, \apj, 
632, 859 

\bibitem[Mickaelian {\etal}(2006)]{mickaelian06} Mickaelian, A.~M.,
Hovhannisyan, L.~R., Engels, D., Hagen, H.-J., \& Voges, W.\ 2006,
\aap, 449, 425

\bibitem[Moster {\etal}(2010)]{moster10} Moster, B.~P., Somerville,
R.~S., Maulbetsch, C., {\etal} 2010, \apj, 710, 903

\bibitem[Mullis(2001)]{mullis01} Mullis, C.~R.\ 2001, Ph.D.~Thesis,
IfA, University of Hawaii


\bibitem[Naab {\etal}(2007)]{naab07} Naab, T., Johansson, P.~H.,
Ostriker, J.~P., \& Efstathiou, G.\ 2007, \apj, 658, 710

\bibitem[Naab {\etal}(2009)]{naab09} Naab, T., Johansson, P.~H., \&
Ostriker, J.~P.\ 2009, \apjl, 699, L178

\bibitem[Osterbrock \& Ferland(2006)]{agnsquared} Osterbrock, D. E.,
\& Ferland, G. J. 2006, Astrophysics of Gaseous Nebulae and Active
Galactice Nuclei University Science Books, ISBN: 1-891389-34-3

\bibitem[Rauch(1998)]{rauch98} Rauch, M.\ 1998, \araa, 36, 267

\bibitem[Rees \& Ostriker(1977)]{rees77} Rees, M.~J., \& Ostriker,
J.~P.\ 1977, \mnras, 179, 541

\bibitem[Ribaudo {\etal}(2011)]{ribaudo11} Ribaudo, J., Lehner, 
N., Howk, J.~C., et al.\ 2011, \apj, 743, 207 

\bibitem[Savage \& Sembach(1991)]{savage91} Savage, B.~D., \& Sembach,
K.~R.\ 1991, \apj, 379, 245

\bibitem[Schaye(2001)]{schaye01} Schaye, J.\ 2001, \apj, 559, 507

\bibitem[Schneider {\etal}(1993)]{schneider93} Schneider, D.~P.,
{\etal} 1993, ApJS, 87, 45

\bibitem[Schlegel {\etal}(1998)]{schlegel98} Schlegel, D.~J., 
Finkbeiner, D.~P., \& Davis, M.\ 1998, \apj, 500, 525 

\bibitem[Sembach {\etal}(2003)]{sembach03} Sembach, K.~R., Wakker,
B.~P., Savage, B.~D., {\etal} 2003, \apjs, 146, 165

\bibitem[Shaw {\etal}(2009)]{cos-dhb} Shaw, B. {\etal} 2009, COS Data
Handbook, Version 1.0, (Baltimore: STScI).

\bibitem[Shull {\etal}(1998)]{shull98} Shull, J.~M., Penton, S.~V.,
Stocke, J.~T., {\etal} 1998, \aj, 116, 2094

\bibitem[Shull {\etal}(2003)]{shull03} Shull, J.~M., Tumlinson, J., \&
Giroux, M.~L. 2003, \apjl, 594, L107

\bibitem[Silk(1977)]{silk77} Silk, J.\ 1977, \apj, 211, 638

\bibitem[Simard {\etal}(2002)]{simard02} Simard, L., Willmer,
C. N. A., Vogt, N. P., Sarajedini, V. L., Philips, A. C., Weiner,
B. J., Koo, D. C., Im, M., Illingworth, G. D., \& Faber, S. M. 2002,
ApJS, 142, 1

\bibitem[Simcoe {\etal}(2004)]{simcoe04} Simcoe, R.~A., Sargent, 
W.~L.~W., \& Rauch, M.\ 2004, \apj, 606, 92

\bibitem[Steidel {\etal}(1994)Steidel, Dickinson, \& Persson]{sdp94} Steidel, C.~C., 
Dickinson, M., \& Persson, S.~E.\ 1994, \apjl, 437, L75 

\bibitem[Steidel \& Sargent(1992)]{ss92} Steidel, C.~C., \& Sargent,
W.~L.~W.\ 1992, \apjs, 80, 1

\bibitem[Stewart(2011)]{stewart11} Stewart, K. R. 2011, arXiv:1109.3207v1

\bibitem[Stewart(2012)]{stewart12} Stewart, K. R. 2012, private
communication

\bibitem[Stewart {\etal}(2009)]{stewart09} Stewart, K.~R., Bullock,
J.~S., Wechsler, R.~H., \& Maller, A.~H.\ 2009, \apj, 702, 307

\bibitem[Strader {\etal}(2011)]{strader11} Strader, J., Romanowsky,
A.~J., Brodie, J.~P., {\etal} 2011, \apjs, 197, 33

\bibitem[Strutskie {\etal}(2006)]{strutskie06} Strutskie, M. F.,
Cutri, R. M., Stiening, R. {\etal} 2006, AJ, 131, 1163

\bibitem[Sutherland \& Dopita(1993)]{mappings} Sutherland, R.~S., \&
Dopita, M.~A.\ 1993, \apjs, 88, 253

\bibitem[Swindle {\etal}(2011)]{swindle11} Swindle, R., Gal, R.~R., La
Barbera, F., \& de Carvalho, R.~R.\ 2011, \aj, 142, 118

\bibitem[Thom {\etal}(2011)]{thom11} Thom, C., Werk, J.~K., Tumlinson,
J., et al.\ 2011, \apj, 736, 1

\bibitem[Tripp {\etal}(2001)]{tripp01} Tripp, T.~M., Giroux, M.~L.,
Stocke, J.~T., Tumlinson, J., \& Oegerle, W.~R.\ 2001, \apj, 563, 724

\bibitem[Trujillo-Gomez {\etal}(2011)]{trujillo11} Trujillo-Gomez, 
S., Klypin, A., Primack, J., \& Romanowsky, A.~J.\ 2011, \apj, 742, 16 


\bibitem[van de Voort {\etal}(2011)]{vandevoort11} van de Voort, F.,
Schaye, J., Booth, C.~M., Haas, M.~R., \& Dalla Vecchia, C.\ 2011,
\mnras, 414, 2458

\bibitem[van de Voort \& Schaye(2011)]{vandevoort+scahye11} van de
Voort, F., \& Schaye, J.\ 2011, arXiv:1111.5039

\bibitem[van der Wel et al.(2005)]{vanderwel05} van der Wel, A.,
Franx, M., van Dokkum, P.~G., et al.\ 2005, \apj, 631, 145

\bibitem[Verner \& Iakovlev(1990)]{verner90} Verner, D.~A.,
\& Iakovlev, D.~G.\ 1990, \apss, 165, 27

\bibitem[Voges {\etal}(1999)]{voges99} Voges, W.,
Aschenbach, B., Boller, T., et al.\ 1999, \aap, 349, 389

\bibitem[Weymann {\etal}(1991)]{weymann91} Weymann, R.~J., Morris,
S.~L., Foltz, C.~B., \& Hewett, P.~C.\ 1991, \apj, 373, 23
\bibitem[Haardt \& Madau(2011)]{haardt11} Haardt, F., \& Madau, P.\
2011, ApJ, arXiv:1105.2039

\bibitem[White \& Rees(1978)]{white78} White, S.~D.~M., \& Rees,
M.~J.\ 1978, \mnras, 183, 341

\end{thebibliography}
\end{document}